\documentclass{revtex4-1}
\usepackage{graphicx}
\usepackage{dcolumn}
\usepackage{bm}
\usepackage[colorlinks=true, linkcolor=blue, citecolor=red]{hyperref}

\usepackage{amsmath}
\usepackage{amssymb}
\usepackage{amsfonts}
\usepackage{resizegather}
\usepackage{caption}
\usepackage{color}
\usepackage{subfigure}
\usepackage{ulem}
\usepackage[para,online,flushleft]{threeparttable}

\begin{document}
\preprint{APS/123-QED}
\title{Fourier-series expansion of the dark-energy equation of state}

\author{David Tamayo}
\email{dtamayo@fis.cinvestav.mx}
\affiliation{Departamento de F\'isica, Centro de Investigaci\'on y de Estudios Avanzados del IPN, A.P. 14-740,
07000 M\'exico D.F., M\'exico.}
\author{J. Alberto Vázquez}
\email{javazquez@icf.unam.mx}
\affiliation{Instituto de Ciencias Fisicas, Universidad Nacional Autonoma de Mexico, Apdo. Postal 48-3, 62251 Cuernavaca, Morelos, Mexico}

\begin{abstract}
The dark-energy component of the Universe still remains a mystery; however, several papers based on observational data have shown that its equation of state may have an oscillatory behaviour. 
In this paper, we provide a general description for the dark-energy equation of state $w(z)$ in the form of a Fourier series. 
This description generalizes some previous dynamical dark-energy models and is in agreement with the $w(z)$ reconstructions. 
We make use of a modified version of a simple and fast Markov chain Monte Carlo code to constrain the model parameters. 
For the analysis we use data from supernovae type Ia, baryon acoustic oscillations, $H(z)$ measurements and cosmic microwave background. 
We provide a comparison of the proposed model with $\Lambda$CDM, $w$CDM and the standard Taylor approximation. 
The Fourier-series expansion of $w(z)$ is preferred from $\Lambda$CDM at more than the 3$\sigma$ significance level based on the improvement in the fit alone. 
We use the Akaike criterion to perform the model comparison and find that, even though there are extra parameters, there is a slight preference for the Fourier series compared with the $\Lambda$CDM model. 
The preferred shape of $w(z)$ found here puts in jeopardy the single scalar field models, as they cannot reproduce the crossing of
the phantom divide line $w = −1$.
\end{abstract}


\maketitle

\section{Introduction}

The dark energy is still an unknown negative pressure cosmic component for which the 
simplest case is given in terms of a perfect fluid with an equation of state (EoS) 
$p=w\rho$ with $w=-1$, this particular model is commonly named as the cosmological 
constant $\Lambda$ and it is a key piece of the standard cosmological model: the
$\Lambda$CDM model.
Even though the standard cosmological model fits well with most of the current 
astronomical observations, there exist important tensions among different recent data sets.
For instance, the value of $H_0$ measured from CMB data by the Planck Collaboration \cite{Ade:2015xua} is 
3.4$\sigma$ lower than the local value reported by Riess et al. \cite{Riess:2016jrr}.
The matter density fraction consistent with the Lyman-$\alpha$ forest measurement of the baryon 
acoustic oscillations (BAO) is smaller than the one preferred by CMB measurements \cite{Delubac:2014aqe}.
On the other hand, based on the most minimal a priori assumptions, model independent 
reconstructions of the evolution of the dark energy EoS parameter  
exhibit a dynamical behaviour of  $w(z)$ \cite{Vazquez:2012ce, Wang:2018fng, Zhao:2017cud, Hee:2016nho}, 
putting in tension the $\Lambda$CDM model amongst several models for which $w=constant$.
From the theoretical point of view, the standard cosmological model also carries out several 
important theoretical problems such as the absence of physical grounds to justify the 
cosmological constant, the coincidence problem and fine tuning, see 
\cite{Padilla:2015aaa, Velten:2014nra,Lopez-Corredoira:2017rqn}.
In order to get around these issues, there have been plenty of proposals to describe 
the general behaviour of this dark component, i.e. scalar fields (quintessence, K-essence, 
phantom, quintom, non-minimally coupled scalar fields, etc.) \cite{Arun:2017}, 
modified gravity \cite{Clifton:2011jh}, interacting dark energy \cite{Zimdahl:2012mj} 
and divergence free parameterizations \cite{Akarsu:2015yea}, among many others \cite{Yoo:2012ug}.

Dark energy parameterizations of the EoS parameter are a set of phenomenological models that 
consist of assuming that the dark energy behaves as a perfect fluid with a dynamical equation 
of state, that is $p=w(z)\rho$, without any other assumption about the origin of this 
behaviour from fundamental physics.
The main goal of this approach is to model the time evolution of $w(z(t))$ from observational 
data which can give important insights about the evolution of the dark energy and put a 
phenomenological basis for its theoretical description.
Here we briefly summarise some of these models. 

First, one of the most popular time-evolving parameterizations consists of expanding 
$w$ in a Taylor series around $a=1$ (today): $w(a) = \sum_{i=0}^N (1-a)^i w_i$, where $N$ defines 
the order of the polynomial expansion and $w_i$ are constant values.
The trivial case, when $N=0$, is $w(a)=w_0$ also known as $w$CDM model, and clearly for the particular case of $w_0=-1$ 
we return to the well-known cosmological constant. 
The case $N=1$, i.e. $w(a)=w_0 +(1-a)w_1$ corresponds to the very well-known 
Chevallier-Polarski-Linder (CPL) model \cite{Chevallier:2000qy, Linder:2002et}, this 
richer model has been widely studied \cite{Kumar:2012gr} and can be mapped into 
another dark energy models like quintessence and barotropic dark energy \cite{Scherrer:2015tra}.
A recent study of several cases of $w$CDM and CPL models can be found in \cite{Sagredo:2018rvc}, 
here the authors compare the models by performing different statistical criteria in order 
to highlight the differences.
For higher orders $N=2, 3, 4$, a nice analysis is performed in \cite{Dai:2018}, the authors 
by analysing cosmological data (SNIa, CMB, LSS, $H(z)$, BAO) conclude that the 
concordance cosmological constant model ($w=-1$) is still safely consistent with 
these observational data at the 68$\%$ confidence level. However, when adding the 
high-redshift BAO measurement from the Ly$\alpha$ forest of BOSS DR11 quasars into 
the calculation there is a significant impact on the reconstruction result, the $w$ 
prefers values significantly smaller than -1.
\\

Parameterizations which use power laws and exponential functions with two and three 
free parameters were analysed by Martins et al. in \cite{Martins:2018bzo}. 
Through a standard statistical analysis of cosmological data (SNIa, $H(z)$) they set 
constraints at the present-day values of the dark energy EoS and in the asymptotic 
past in these models.
They conclude that the dark energy EoS near the present day must be very similar to 
that of a cosmological constant, and any significant deviations from this behaviour 
can only occur in the deep matter era. 

Another approach consists in phenomenological models using trigonometric functions.
A first attempt was proposed by Linder arguing that oscillating dark energy models 
offer one idea for solving the coincidence problem \cite{Linder:2005dw}, the ansatz 
considers an EoS parameter with the form of $w(a)=w_0 -A \sin(B\ln a)$ where the 
natural period of the cosmic expansion is given by $H^{-1} = (d \ln a/dt)^{-1}$, 
so they examine periodicity in units of the $e$-folding scale $\ln a$.
A further work of Pace et al. \cite{Pace:2011kb} studies the imprints on the formation 
and evolution of cosmic structures in six different variations of this particular 
dark energy model. 
Moreover, from a phantom scalar field conformally coupled to gravity model 
Kurek et al. obtain a parameterization of the type 
$w(z)= -1 +(1+z)^3[ C_1\cos(\ln(1 + z)) + C_2 \sin(\ln(1 +z))]$ \cite{Kurek:2007bu}. 
In the reference \cite{Pan:2017zoh} Pan et al. study several oscillating dark energy models: 
$w_I = w_0 +b[1-\cos(\ln(1+z))]$, $w_{II} = w_0 +b\sin(\ln(1+z))$, 
$w_{III}=w_0 +b[\frac{\sin(1+z)}{1+z} - \sin1]$ and $w_{IV}=w_0 +b[\frac{z}{1+z}\cos(1+z)]$, 
see references therein to see the motivation of the selection of these particular 
phenomenological models. 
They perform a confrontation of these dark energy parameterizations with observational 
data (JLA, SNIa, BAO, CMB,  redshift space distortion, weak gravitational lensing, $H(z)$) 
they find that the best-fit characters of almost all models are bent towards 
the phantom region; nevertheless, in all of them, the quintessential regime is 
also allowed within $1\sigma$ confidence level. 
Finally, they perform the Bayesian analysis, which shows that the current 
observational data supports the $\Lambda$CDM paradigm over this set of 
oscillating dark energy parameterizations.
A complementary study for this set of models is done in \cite{Panotopoulos:2018sso}, 
where the authors compute the statefinder parameters, and at the level of linear 
cosmological perturbations they compute the growth index as well as the combination 
parameter $f\sigma_8$, obtaining similar conclusions.
In Jaime et al. \cite{Jaime:2018ftn} the authors propose a cosine-like parameterization 
for the dark energy EoS with the form $w= -1 +\frac{w_0}{1+w_1\,z^{w_2}}\cos(w_3+z)$ 
which can reproduce some successful $f(R)$ gravity models with a precision 
between [0.5$\%$-0.8$\%$] over the numerical solutions, using observational data 
from BAO, SNIa, and cosmic chronometers they investigate the constraints on the new EoS parameters.

The aforementioned reconstruction of $w(z)$ from observational data shows that the 
shape of $w(z)$ crosses the phantom divide line (PDL) $w=-1$ several times, having an 
oscillating behaviour and puts in jeopardy many dark energy models such as $\Lambda$CDM 
and single scalar field models that are unable to cross the PDL.
This oscillating $w(z)$ can be modelled with Taylor series of third order or more, 
but has its limitations; first, the expansion has to be done around a specific point 
which commonly is around $z=0$ and this implies that far away from the expansion point the 
approximation may not be accurate; second, the expansion is done with polynomials and to describe 
an oscillatory function many terms are needed to get a  good representation and third, 
sufficiently far from $z=0$ the polynomial always grows or decreases monotonically.

If we want to model an oscillating $w(z)$, trigonometric functions are a natural 
choice and therefore in this work we generalise this idea using a general description of $w(z)$ 
in the form of Fourier series. This approach avoids the Taylor expansion problems 
mentioned before: the expansion is done over a period and not a point, there are 
needed less terms of the series to reproduce well the oscillations and the 
trigonometric functions are bounded.
We show that the Fourier series approach fits better to the data than 
Taylor expansion models. 
\\

The present work is organised in the following way: first in section (\ref{section: The Model}) 
we provide a general description for the model proposed in this work. Then in section 
(\ref{section: Methology}) the methodology of the data analysis is done.
The results and constraints of the proposed models as well as its comparison with 
different models are shown in (\ref{section: Results}). 
Finally the conclusion and discussion of the results are presented in (\ref{section: Conclusions}).

\section{The oscillating dark energy model}\label{section: The Model}

\begin{figure}
 \includegraphics[trim = 0mm  0mm 0mm 0mm, clip, width=5.4cm]{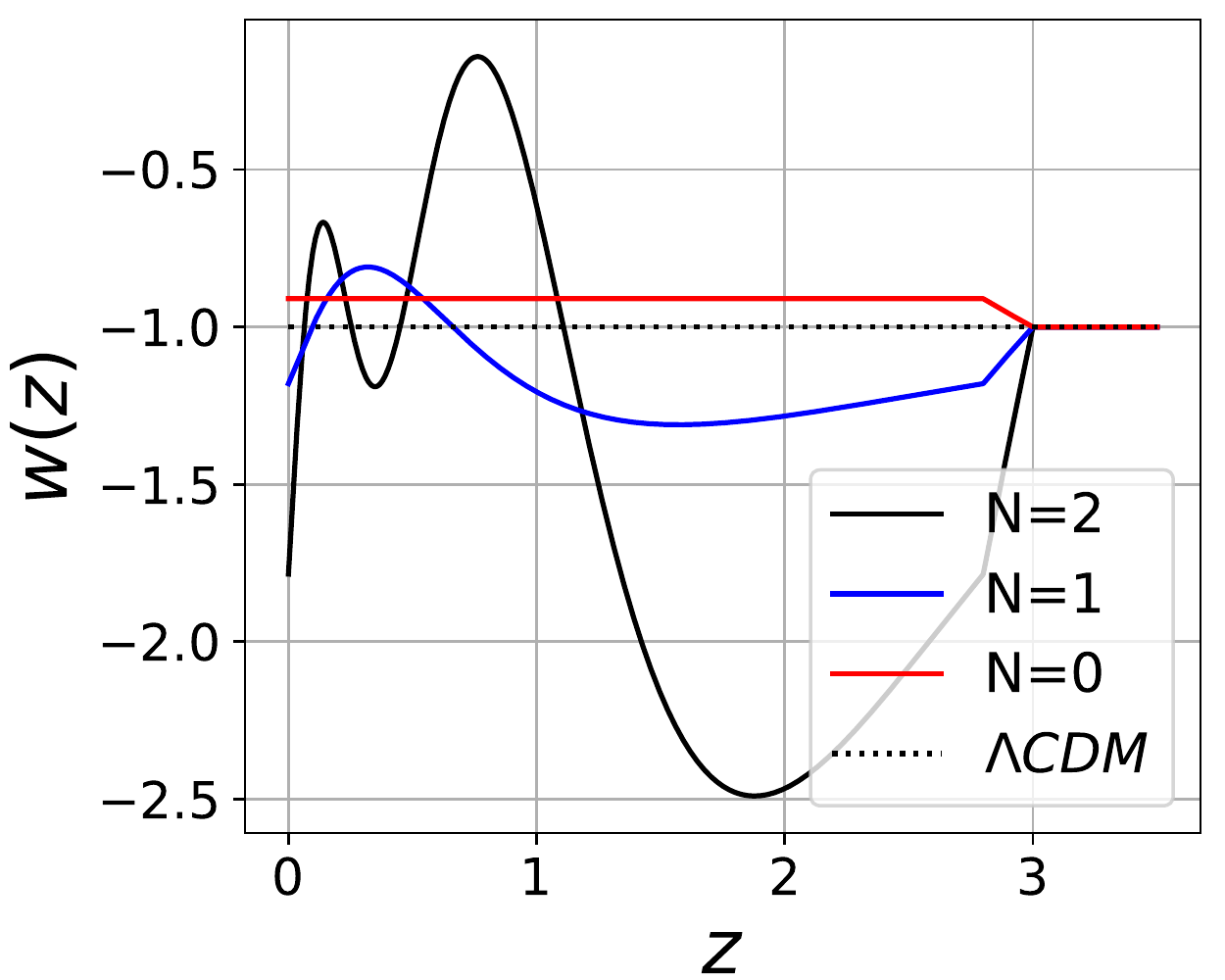}
  \includegraphics[trim = 0mm  0mm 0mm 0mm, clip, width=5.4cm]{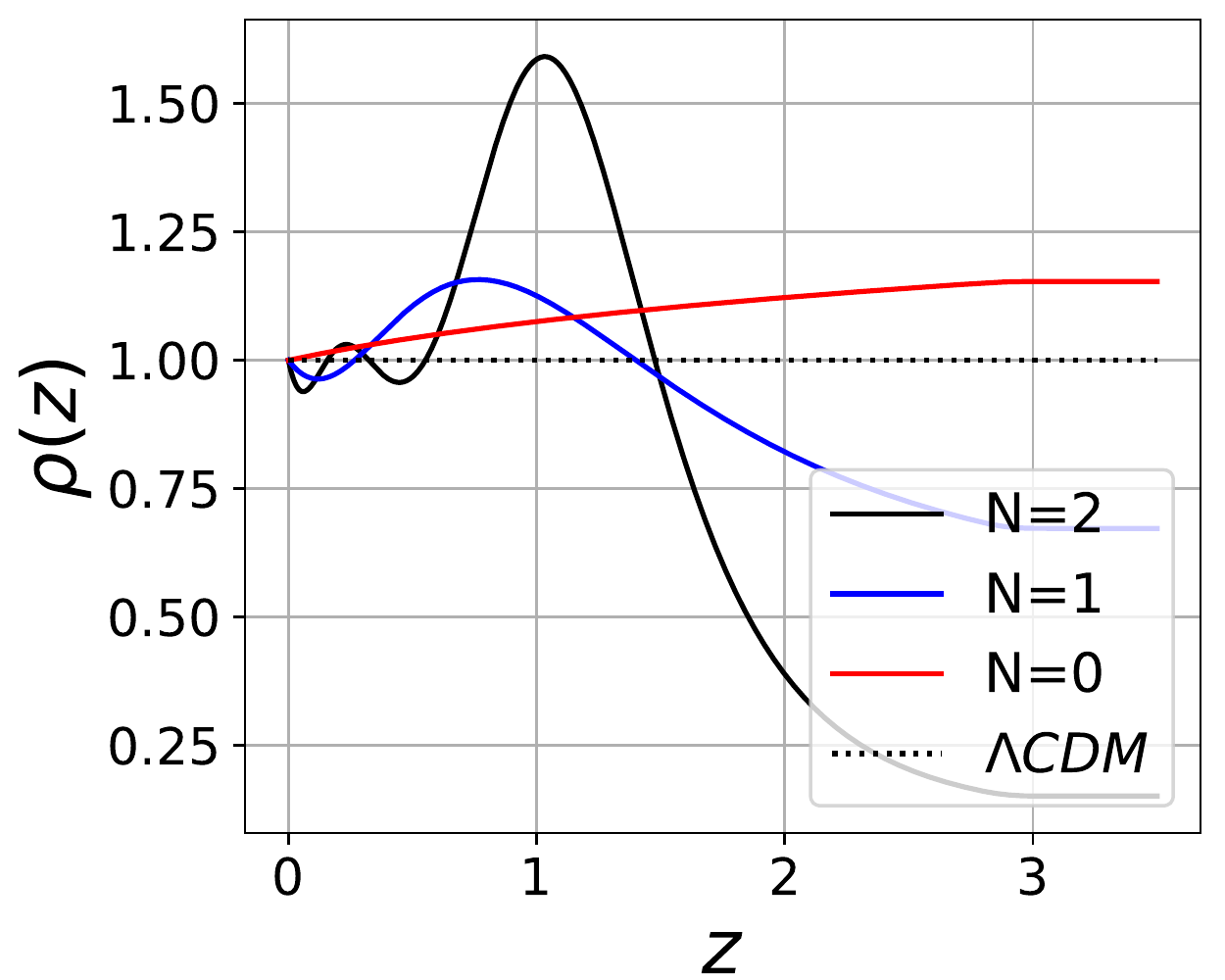}
  \includegraphics[trim = 0mm  0mm 0mm 0mm, clip, width=5.4cm]{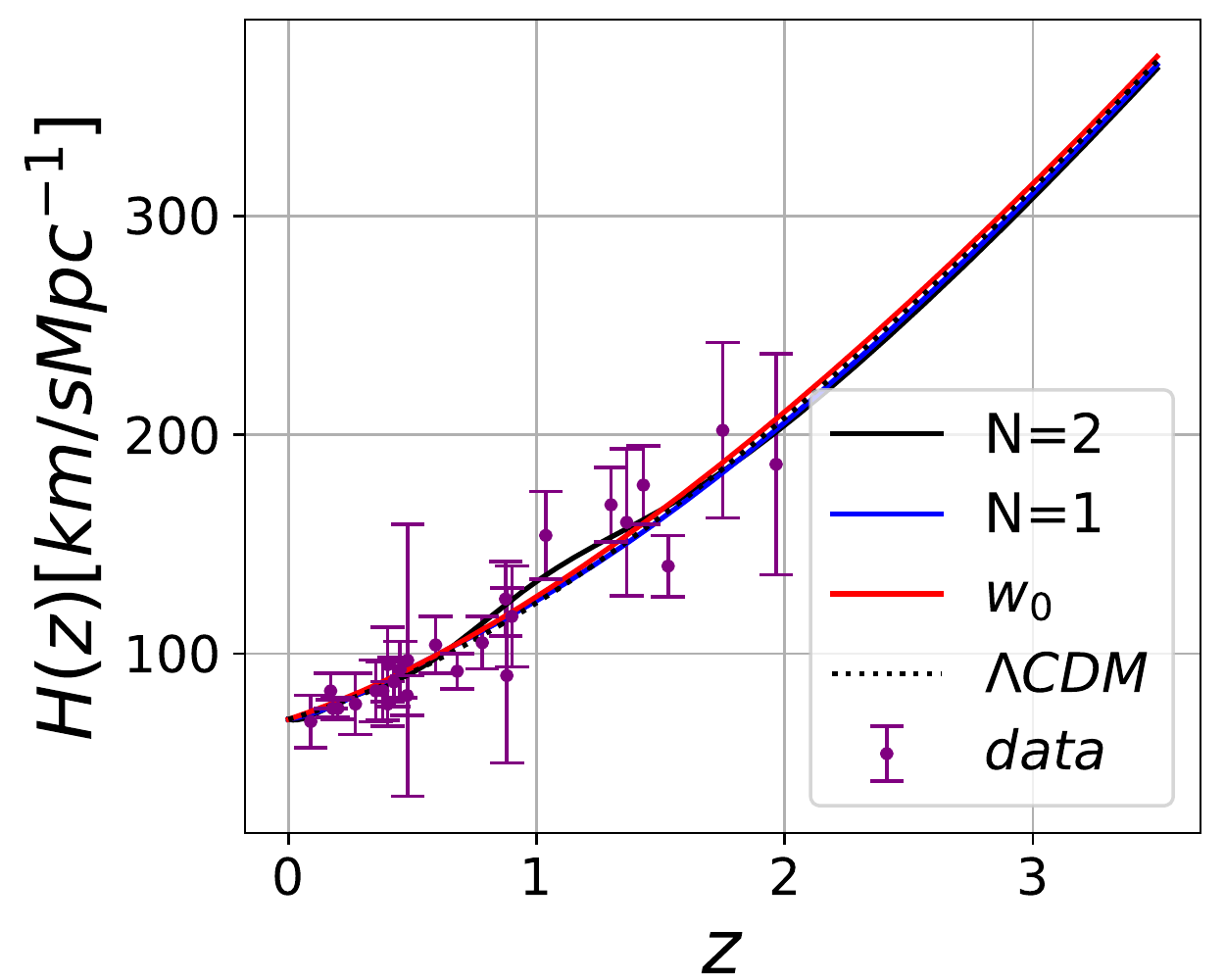}
\caption{{\bf Left panel}: dark energy EoS parameter $w(z)$, {\bf middle panel}: energy density $\rho(z)$ of the dark energy and {\bf right panel}: Hubble function $H(z)$ with data from \cite{Gomez-Valent:2018hwc}. In solid black and blue lines the first two harmonics of the Fourier series $N=2,1$ (number of pairs of $a_n$ and $b_n$ of the expansion), in red the case with only the constant term $w_0$ of the series and the dotted black line corresponds to the $\Lambda$CDM model. The used values of $a_n$, $b_n$ and $w_0$ for this plot were taken from Table \ref{table}.}
\label{fig:rho}
\end{figure}

The general idea of this work is to consider the EoS parameter $w(a)$ as a Fourier series 
in the interval $1 \geq a \geq a_{med}$ (remembering the convention of the value of the 
scale factor at present time $a(t_0) \equiv 1$), a linear adjust in the interval 
$a_{med} \geq a \geq a_{ini}$ and $w=-1$ otherwise. 
The motivation of selecting these intervals is the following. 
First, the available data shows a possible oscillating behaviour of $w$ at low 
redshifts ($0\leq z \lesssim 2.5$) \cite{Zhao:2017cud}, thus the first natural 
approximation is to use trigonometric functions and then in a more general form 
use the Fourier series. 
Then, at higher redshifts ($z\gtrsim 3$) there is not enough evidence about the 
behaviour of $w$ different from the cosmological constant, i.e. we use $w=-1$.
To join these two different behaviours in a continuous way we use a transitional 
interval in which $w$ is linear.
We impose that at the transition points, $a_{med}$ and $a_{ini}$, the value of the functions used by our model is continuous, $w_{F}(a_{med})=w_{L}(a_{med})$ and $w_{L}(a_{ini})= w_{C} =-1$, where the subscripts $F,L,C$ correspond to the Fourier, linear and constant 
parts of $w$.
Therefore, the piecewise function of the EoS parameter $w$ in terms of the scale 
factor is: 
\begin{equation}\label{eq:w_Fourier}
w(a) = \left\{
        \begin{array}{ll}
            -1 & \quad a_{ini} > a,  \\
            m\,(a-a_{ini}) + b & \quad a_{med} \geq a \geq a_{ini}, \\
            {\displaystyle \frac{w_0}{2}+\sum_{n=1}^{\infty}\left(a_{n}\sin \left[{\frac {2n\pi }{T}}(a-a_{med})\right] + b_{n}\cos \left[{\frac {2n\pi }{T}}(a-a_{med})\right]\right)} & \quad 1 \geq a \geq a_{med}.
        \end{array}
    \right.
\end{equation}
In terms of the redshift $a=(z+1)^{-1}$ we have
\begin{equation}\label{eq:wz_Fourier}
w(z) = \left\{
        \begin{array}{ll}
            -1 & \quad z_{ini} < z, \\
            m\,\left(\frac{1}{z+1} -\frac{1}{z_{ini}+1}\right) + b & \quad z_{med} \leq z \leq z_{ini}, \\
            \frac{w_0}{2} +\sum_{n=1}^{\infty} \left(a_n\sin \left[\frac{2n\pi}{T} \left(\frac{1}{z+1} -\frac{1}{z_{med}+1}\right)\right] +b_n\cos \left[\frac{2n\pi}{T} \left(\frac{1}{z+1} -\frac{1}{z_{med}+1}\right)\right] \right) & \quad 0 \leq z \leq z_{med}.
        \end{array}
    \right.
\end{equation}
where $T= 1-a_{med}$ is the period. 
The constants of the linear equation, $m$ and $b$, can be calculated from the equation of a line given two points~($P_1, P_2$): 
\begin{eqnarray}
P_1 &=& (a_{med},\, w(a_{med})) \equiv (a_{med},\, w_{med}) = \left(a_{med},\, \frac{w_0}{2} +\sum_{n=1}^{\infty} b_n\right), \\
P_2 &=& (a_{ini},\, w(a_{ini})) \equiv  (a_{ini},\, w_{ini}) = (a_{med},\, -1),
\end{eqnarray}
therefore 
\begin{eqnarray}
m &=& \frac{w_{med} - w_{ini}}{a_{med} -a_{ini}} = \frac{\frac{w_0}{2} +\sum_{n} b_n +1}{a_{med}-a_{ini}}, \\
b &=& w_{ini} = -1.
\end{eqnarray}
To calculate the dark energy density $\rho_{de}$ we have to solve the conservation equation
\begin{eqnarray}
\dot{\rho}_{de} +3H(1+w)\rho_{de} = 0,\\
\Rightarrow \quad \rho_{de} = A\, \exp\left[-3\int (1+w)\frac{da}{a}\right],
\end{eqnarray}
where $A$ is the integration constant. 
Performing the integral for the different parts of $w(a)$ we have:
\begin{eqnarray}
\rho_F  &=&  A_F\, a^{-3(1 +\frac{w_0}{2})}\, \exp\left[-3\sum_{n=1}^{\infty}\operatorname{Ci}(a\,n\theta) [b_n \cos(a_{med}n\theta) -a_n \sin(a_{med}n\theta)] \right]\times \nonumber\\
&{}& \exp \left[-3\sum_{n=1}^{\infty}\operatorname{Si}(a\,n\theta) [a_n \cos(a_{med}n\theta) +b_n \sin(a_{med}n\theta)] \right],\\  
\rho_L &=&  A_L\, a^{1+b+3m\,a_{ini}} e^{3m\,a},\\
\rho_C &=& A_C.
\end{eqnarray}
Where $A_F, A_L, A_C$ are the integration constants for each particular case, $\theta = 2\pi/T$ and, $\operatorname{Ci}$ and $\operatorname{Si}$ are the cosine integral and sine integral defined as:
\begin{eqnarray}
\operatorname{Ci}(x) = -\int_x^\infty {\frac {\cos t}{t}}\,dt, \quad
\operatorname{Si}(x) = \int_0^x{\frac {\sin t}{t}}\,dt,
\end{eqnarray}
which in the interval $1 \geq a \geq a_{med}$ are well defined.
The integration constants are calculated by solving the system: 
$\rho_F(a_{med}) = \rho_L(a_{med})$, $\rho_L(a_{ini}) = \rho_C(a_{ini})$ and to close the system we use $\rho_F(a(t_0)) \equiv \rho_0$ (the energy density of the dark energy today). 
Solving, we obtain the integration constants:
\begin{eqnarray}
A_F &=& \rho_{de}^{(0)}  \exp\left[3 \sum_{n=1}^{\infty}\left[\operatorname{Ci}(n\theta) f_n +
\operatorname{Si}(n\theta) g_n\right]\right],\\
A_L &=&  A_F\, a_{med}^{-3(\frac{w_0}{2} - b + a_{ini} m)} \exp\left[
 3 \sum_{n=1}^{\infty} [a_{med} m - f_n \operatorname{Ci}(a_{med}\,n\theta) - g_n \operatorname{Si}(a_{med}\,n\theta)]\right],\\
A_C &=& A_L\, a_{ini}^{-3(1 + b - a_{ini} m)} e^{-3 a_{ini} m},
\end{eqnarray}
where $\rho_{de}^{(0)} = \rho_{de}(a=1) $ and we have defined the auxiliary functions
\begin{eqnarray}
f_n &=& b_n \cos(a_{med}\, n \theta) - a_n \sin(a_{med}\, n \theta),\\
g_n &=& a_n \cos(a_{med}\, n \theta) + b_n \sin(a_{med}\, n \theta).
\end{eqnarray}
Given the energy density of the dark energy, we can calculate the Hubble function for a cosmological model of a universe filled with matter (baryonic and dark matter), radiation (photons, neutrinos, etc.) and dark energy:
\begin{eqnarray}
H = H_0 \left(\Omega_m^{(0)}a^{-3} +\Omega_r^{(0)}a^{-4} +\Omega_{de}^{(0)}\, \rho_{de}/\rho_{de}^{(0)}\right)^{1/2},
\end{eqnarray}
remembering that $\Omega_i=\rho_i/\rho_{crit}$ for the $i-th$ commponent, $\rho_{crit}$ is the critical density of the Universe and $\Omega_i^{(0)}=\rho_i^{(0)}/\rho_{crit}^{(0)} = (8\pi G/3H_0^2)\rho_i^{(0)}$. 
As an example, Figure \ref{fig:rho} displays $w(z)$, $\rho(z)$ and $H(z)$ for the first two harmonics of the Fourier series, the constant case and $\Lambda$CDM.

\section{Methology}\label{section: Methology}

In order to perform the parameter space exploration, and select the best-fit model, 
we make use of a modified version of a simple and fast 
Markov Chain Monte Carlo (MCMC) code that computes expansion
rates and distances from the Friedmann equation, named SimpleMC \cite{Anze:2015, Aubourg:2014yra}. 
The datasets considered throughout the analysis include
a compressed version of the Planck data (PLK), 
a recent reanalysis of SNIa data, and high-precision BAO measurements at different redshifts up to $z<2.36$ \cite{Aubourg:2014yra}.
We also include a collection of currently available $H(z)$ measurements (HD), see \cite{Gomez-Valent:2018hwc} and references therein. 
We assume a flat $\Lambda$CDM universe described by the following parameters:
$\Omega_b h^2$ and  $\Omega_{DM}$ are the physical baryon density and dark matter density, respectively, 
relative to the critical density, and $h$ the dimensionless Hubble parameter 
such that $H_0 = 100h$s$^{-1}$Mpc$^{-1}$km.
Here the neutrinos are massless and the effective number of relativistic species 
has the standard $\Lambda$CDM value of $N_{\rm eff}=3.04$.
The code neglects perturbations for the dark energy, but this could be an important point to 
deal with in a future work.
The SimpleMC code contains the Gelman-Rubin convergence criterion ($R$), which typically is 
set up to be $0.97<R<1.03$;
for an extended review of cosmological parameter inference in cosmology see \cite{Padilla:2019mgi}.
In our Fourier description we introduce a set of free parameters $w_0$, $a_n$, $b_n$
to describe the overall shape of the dark energy equation-of-state $w(z)$.
The transition points selected are $z_{ini}=3.0$ and $z_{med}=2.8$.
For each of them, we allow
variations in amplitudes with conservative flat priors $w_0=[-3, -1]$ and $a_n, b_n = [-1.5, 1.5]$.
Left columns on Table \ref{table} display the parameters used throughout  
each description of $w(z)$.
\\

We perform a model comparison in order to select the best description for the
equation-of-state $w(z)$. 
The main aim of model selection is to balance the goodness of fit to the observational data against the complexity of the model, in this case 
given by the extra free parameters. One way to carry out this process is 
by calculating the Bayesian evidence, which naturally incorporates a
penalisation factor through the prior volume of the parameter space \cite{Vazquez:2012ce, Hee:2016nho}.
Calculating the evidence could be challenging and a computationally 
demanding process. 
In this work, however, for simplicity and noticing the near-gaussianity 
of the posterior distributions, we focus on the information
criteria methods such as the Akaike Information Criterion (AIC).
The AIC is defined as:
\begin{eqnarray}
    {\rm AIC }= -2 \ln \mathcal{L}_{\rm max} + 2k,
\end{eqnarray}
where the first term incorporates the goodness-of-fit through the 
likelihood $\mathcal{L}$, and the second term is interpreted as the 
penalisation factor given by two times the extra number of parameters ($k$) of the model.
The more complex the model is, the faster 
the penalty term takes over. 
For a small number of datapoints $N$, it is important to attach a correction term to 
the AIC \cite{Burnham:2002}, given by

\begin{eqnarray}
    {\rm AIC_C = AIC} + \frac{2k(k+1)}{N-k-1}.
\end{eqnarray}
Therefore the preferred model is the one that minimises the ${\rm AIC_C}$.

\section{Results}\label{section: Results}

\begin{table}
\begin{center}
\caption{Constraints on the set of parameters used on each description for $w(z)$. 
For one-tailed distributions the upper limit
95\% C.L. is given and for two-tailed the 68\% is shown. 
Its corresponding plot is displayed in Figure \ref{fig:posteriors}.
For each parameterization of $w(z)$ we have considered two sets of data: first row contains SNIa+BAO+HD, while the second row additionally includes PLK data. Second column describes the specific model chosen in the text, with $x=\frac{2\pi}{T}(\frac{1}{z+1} - \frac{1}{z_{med+1}})$.}
\begin{threeparttable}
\begin{tabular}{cccccccccc} 
\cline{1-10}\noalign{\smallskip}
\vspace{0.2cm}
& Model &   $w_0$ & $a_1$ & $b_1$ &
     $a_2$ &  $b_2$ &  $-2\ln \mathcal{L}_{\rm max}$ & 
    S/N \tnote{*} &  $\Delta {\rm AIC}_{C}$  \\
 \hline
\vspace{0.1cm}
&$\Lambda$CDM&  $-2$ & 0 & 0 & 0 & 0 & 68.98	& 0 & 0 	\\
&   &  $-2$ & 0 & 0 & 0 & 0 & 73.44	& 0 & 0 	\\
	    	
\hline
\vspace{0.1cm}
(a)&$\frac{w_0}{2}$	
& $-1.82\pm 0.12$ & 0 & 0 & 0 & 0  & 67.06	 & $1.4 \sigma$ & $0.3$ \\	
&   & $-1.93\pm 0.08$ & 0 & 0 & 0 & 0  & 73.84	 & $0.5\sigma$  & $2.6$\\
	    	
\hline
\vspace{0.1cm}
(b) &$w_0 + \frac{z}{1+z}a_1$ \tnote{**}	
& $-0.87 \pm 0.11 $& $-0.28\pm 0.58$ & 0 & 0 & 0  & 67.08 & $1.4\sigma$ & $2.7$\\
& & $-0.94 \pm 0.10 $& $-0.11\pm 0.34$ & 0 & 0 & 0  & 73.66 & $0.5\sigma$  & $4.8$\\
	
\hline
\vspace{0.1cm}
(c) & $\frac{w_0}{2}+ a_1\sin(x)$ 
& $-2.06\pm 0.32$ & $-0.16 \pm 0.20$ & 0 & 0 & 0  &	66.74   &  $1.5\sigma$ & $2.3$	\\
&  & $-2.01\pm 0.12$ & $-0.10 \pm 0.11$ & 0 & 0 & 0  &	73.24   &  $0.4\sigma$ & $4.4$	\\
           
\hline
\vspace{0.1cm}
(d) &$\frac{w_0}{2}+ b_1\cos(x)$	
& $-1.78 \pm 0.13$ & 0 & $-0.11\pm 0.12 $& 0 & 0  &	66.12 & $1.7\sigma$  & $1.7$ \\
&  & $-1.94 \pm 0.08$ & 0 & $-0.06\pm 0.11$ & 0 & 0  &  73.50 & $0.2\sigma$  & $4.6$ \\
	        
\hline
\vspace{0.1cm}
(e) & $\frac{w_0}{2}+ a_1\sin(x)$	
& $-2.12\pm 0.33$ & $-0.22\pm 0.21$ & $-0.12\pm 0.13$ & 0 & 0  & 65.84 & $1.8\sigma$ & $3.8$\\
& $+ b_1\cos(x)$ & $-2.09\pm 0.14$ & $-0.19\pm 0.13$ & $-0.17\pm 0.12$ & 0 & 0  & 71.56 & $1.4\sigma$ & $5.1$ \\
        	    
\hline
\vspace{0.1cm}
(f) &$a_1\sin(x) + b_1\cos(x)$ 	
& $-2$ &$-0.09 \pm 0.08$& $-0.54 \pm 0.23$ & $-0.54\pm 0.27$ & 0  & 61.52 & $2.7\sigma$ & $-0.5$\\
&$+ a_2\sin(2x)$ & $-2$ &$-0.07 \pm 0.09$& $-0.26 \pm 0.12$ & $-0.27\pm 0.17$ & 0  & 69.66 & $1.9\sigma$ & $3.2$\\
        	
\hline
\vspace{0.1cm}
(g) &$\frac{w_0}{2} + a_1\sin(x)+ b_1 \cos(x)$	
& $-2.34\pm 0.31$ & $-0.29\pm 0.20$ & $-0.70\pm 0.29$ & $-0.72\pm 0.34$ & 0  & 61.24 & $2.8\sigma$ & $1.7$\\
&$+ a_2\sin(2x)$& $-2.53\pm 0.24$ & $-0.38\pm 0.17$ & $-0.70\pm 0.23$ & $-0.76\pm 0.28$ & 0  & 63.94 & $3.1\sigma$ & $-0.1$\\

\hline
\vspace{0.1cm}
(h)&$\frac{w_0}{2} + a_1\sin(x)+ b_1 \cos(x)$	
& $-2.31\pm 0.31$ & $-0.28\pm 0.22$ & $-0.65\pm 0.30$ & $-0.67\pm 0.34$ & $0.02\pm 0.23$  & 61.40 
& $2.8\sigma$ & $4.4$ \\
&$+ a_2\sin(2x) + b_2\cos(2x)$& $-2.50\pm 0.28$ & $-0.41\pm 0.22$ & $-0.66\pm 0.27$ & $-0.70\pm 0.30$ & $0.13\pm 0.26$  & 64.26 
& $3.0\sigma$ & $0.2$ \\
                        	
\hline
\hline
\end{tabular}
  \begin{tablenotes}
    \item[*] Signal-to-noise ratio = $\sqrt{2 \Delta \ln \mathcal{L}}$ of $w(z)$ deviating from $\Lambda$CDM based on the improvement in the fit alone.
    \item[**] This particular model corresponds to the CPL parameterization.
  \end{tablenotes}
\end{threeparttable}
\label{table}
\end{center}
\end{table}

\begin{figure}
\includegraphics[trim = 0mm 0mm 0mm 0mm, clip, width=10cm, height=6cm]{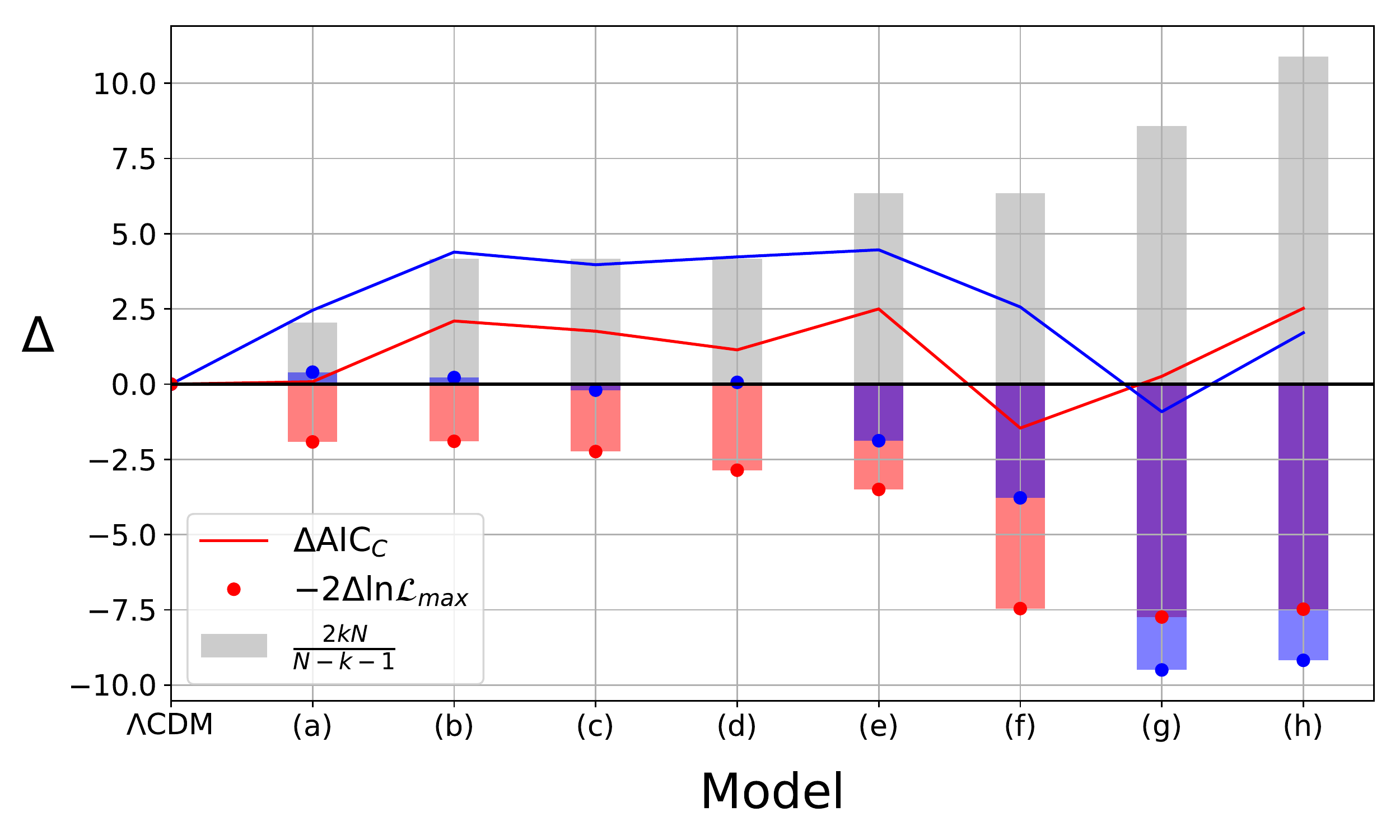}
\caption{{\bf Red bar plots} show the inclusion of extra parameters 
	improves the fit to the data, seen through $-2 \Delta \ln \mathcal{L}_{\rm max}$
	(compare to $\Lambda$CDM). We can go even further adding parameters, 
	however this increment causes that penalty term ({\bf gray bar plot}) 
	dominates and hence the ${\rm AIC_C}$ value raises back again. 
	Therefore, the preferred model is the one that minimises the 
	${\rm AIC_C}$ (solid lines). Blue colour additionally includes PLK data.}
 \label{fig:delta}
\end{figure}

\begin{figure}
\subfigure{            \label{fig:a}
            \includegraphics[trim = 0mm 0mm 0mm 0mm, clip, width=4.2cm, height=3cm]{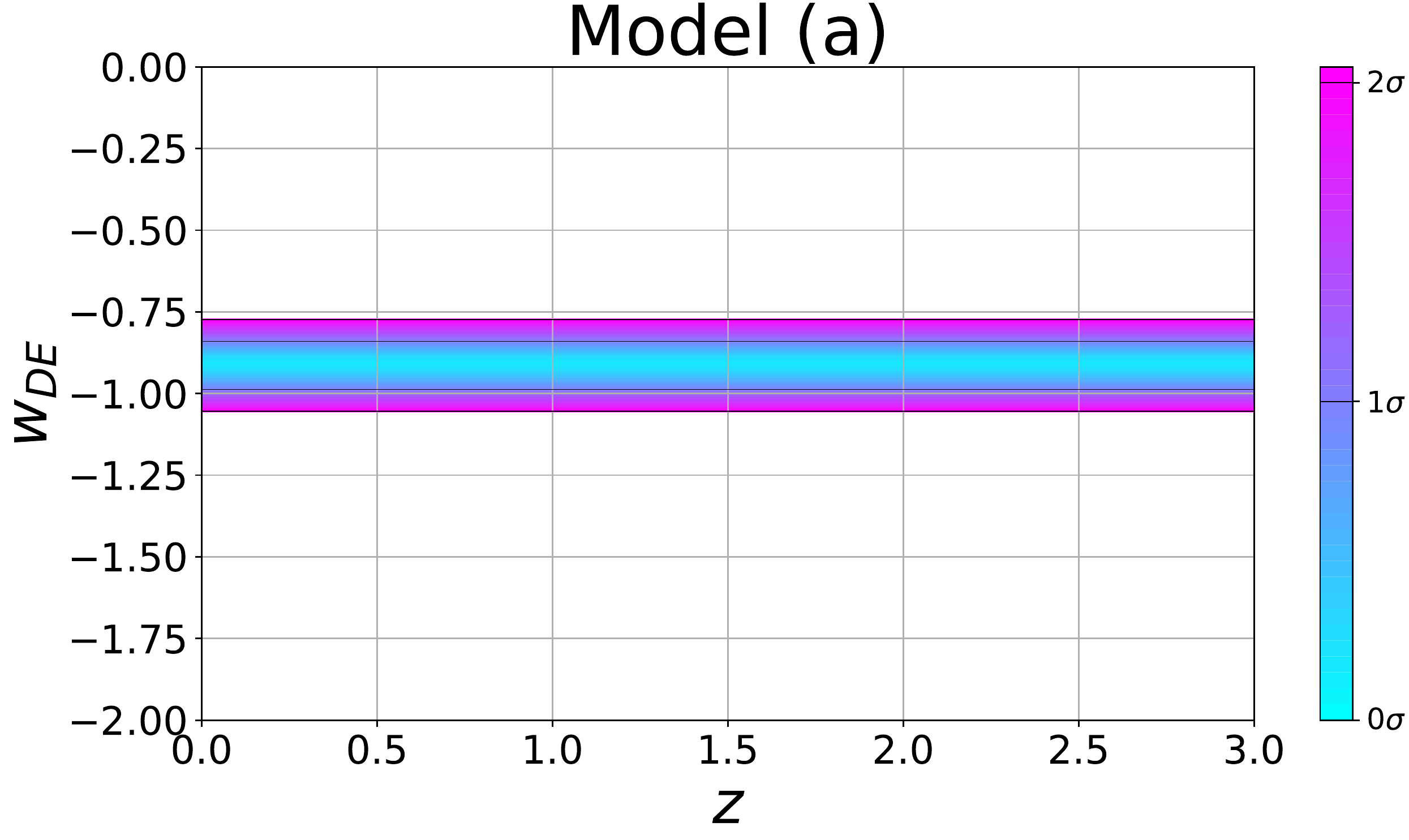}
            \includegraphics[trim = 0mm 0mm 0mm 0mm, clip, width=4.2cm, height=3cm]{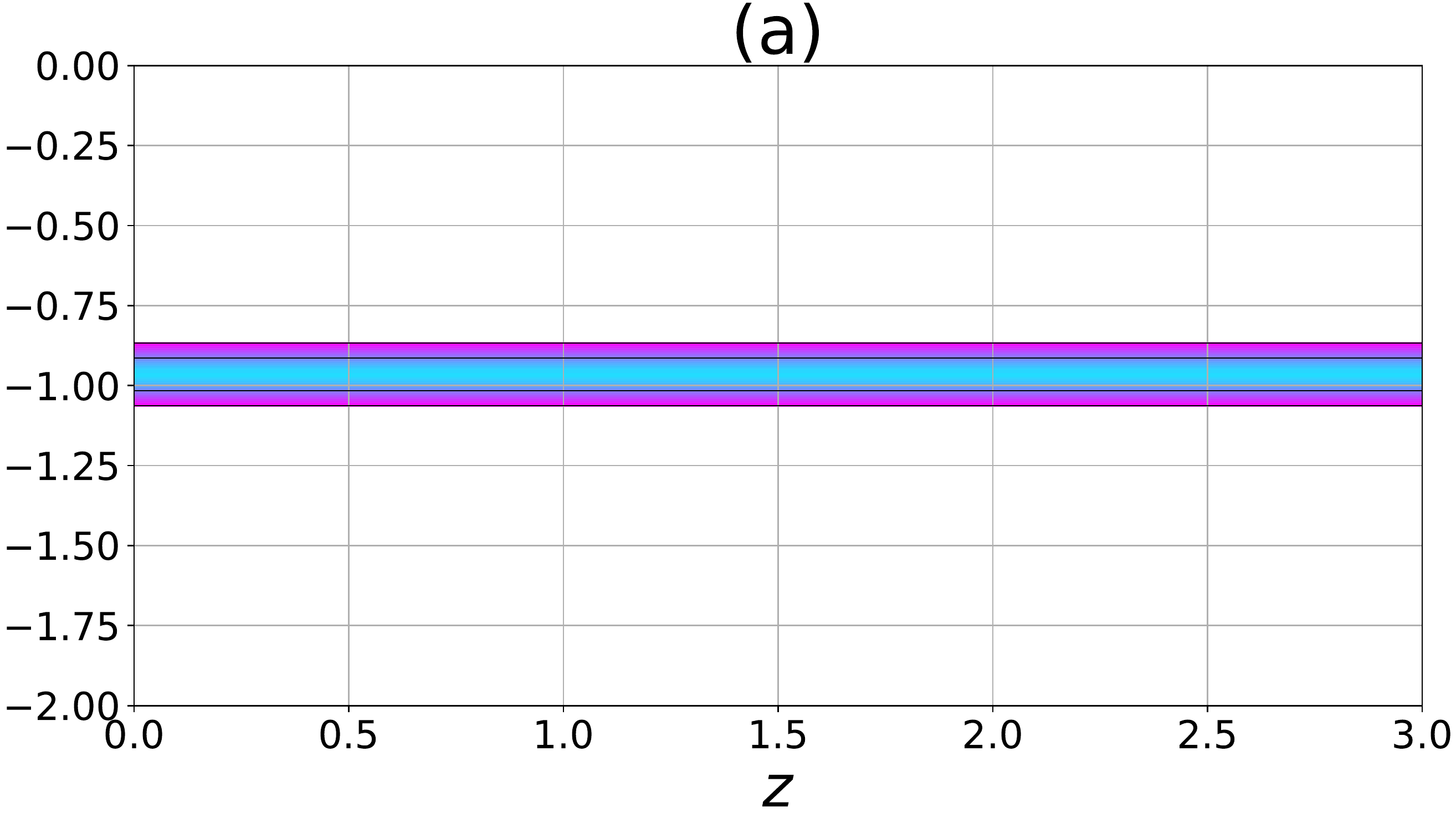}
        }
\subfigure{           \label{fig:b}
          \includegraphics[trim = 0mm 0mm 0mm 0mm, clip, width=4.2cm, height=3cm]{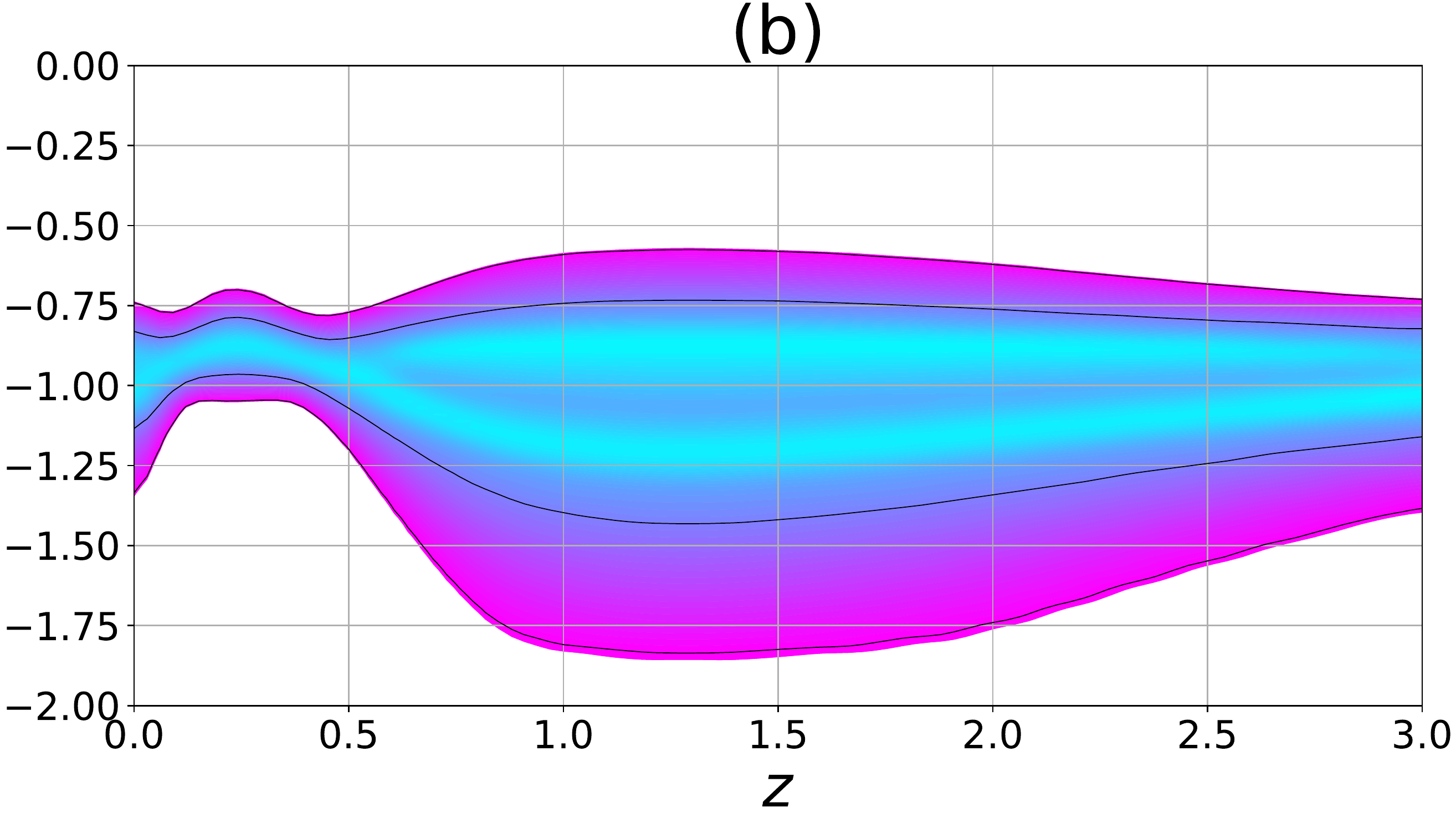}
          \includegraphics[trim = 0mm 0mm 0mm 0mm, clip, width=4.2cm, height=3cm]{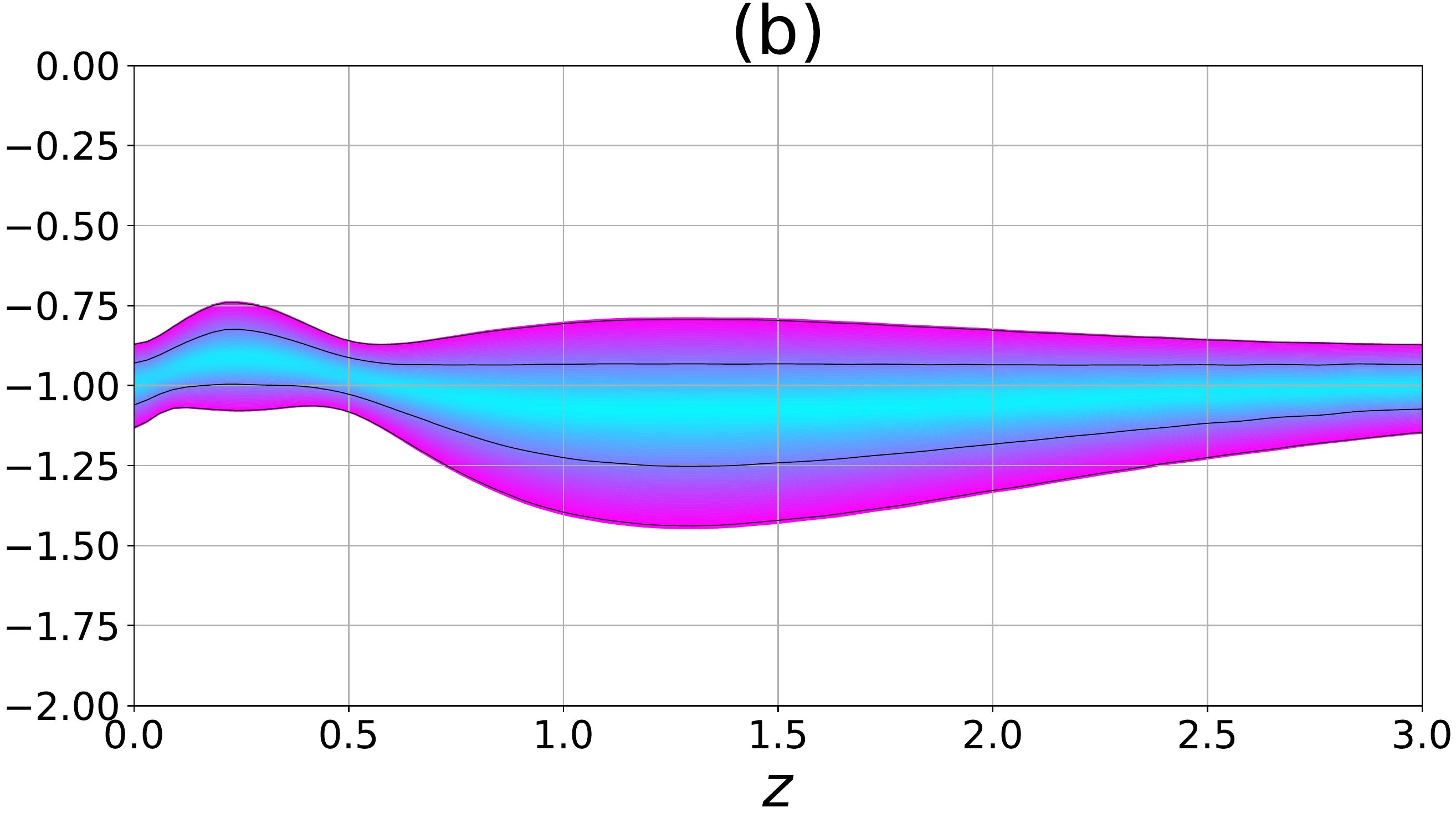}
        }
        
\subfigure{           \label{fig:c}
          \includegraphics[trim = 0mm 0mm 0mm 0mm, clip, width=4.2cm, height=3cm]{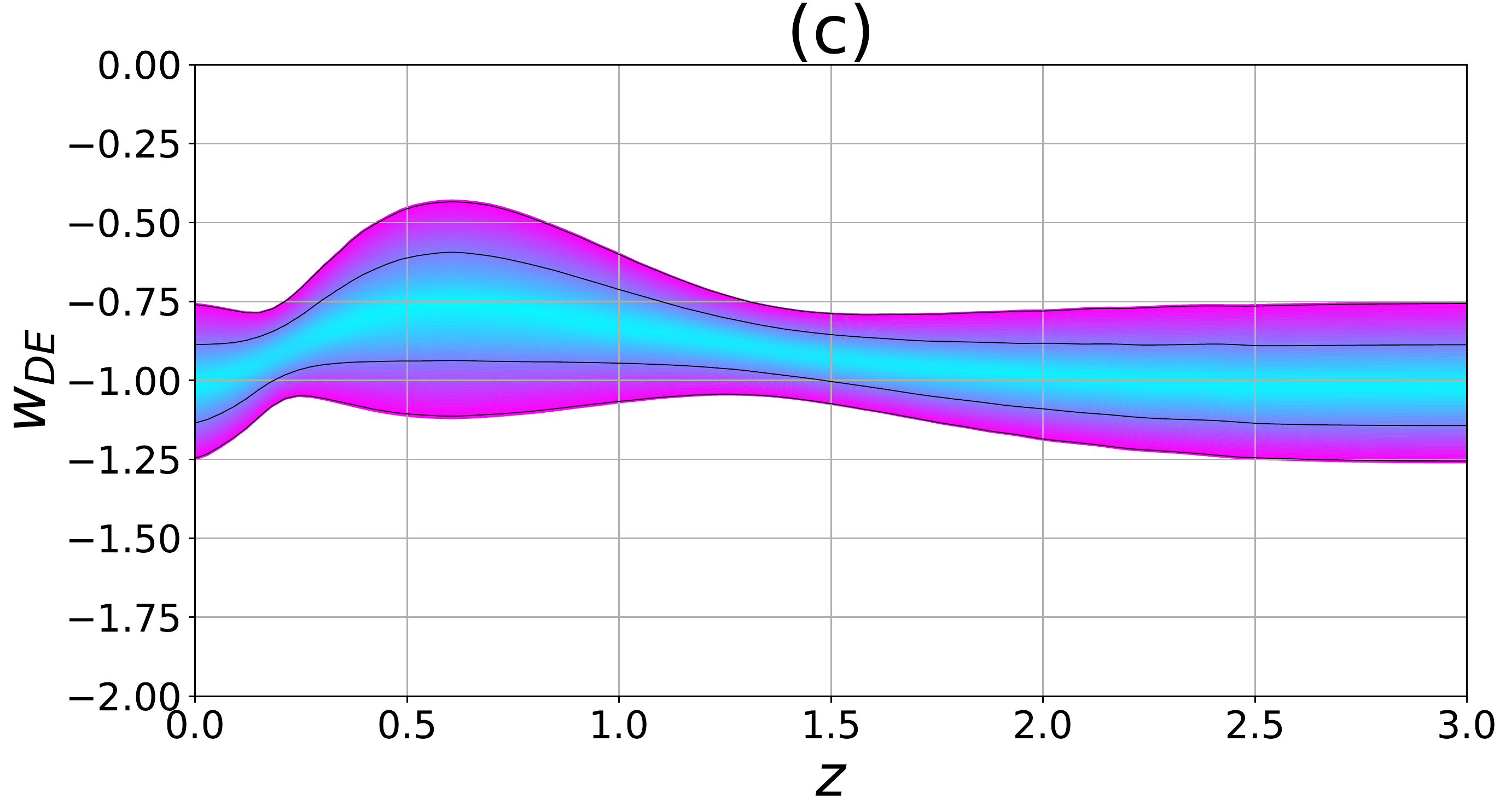}
        \includegraphics[trim = 0mm 0mm 0mm 0mm, clip, width=4.2cm, height=3cm]{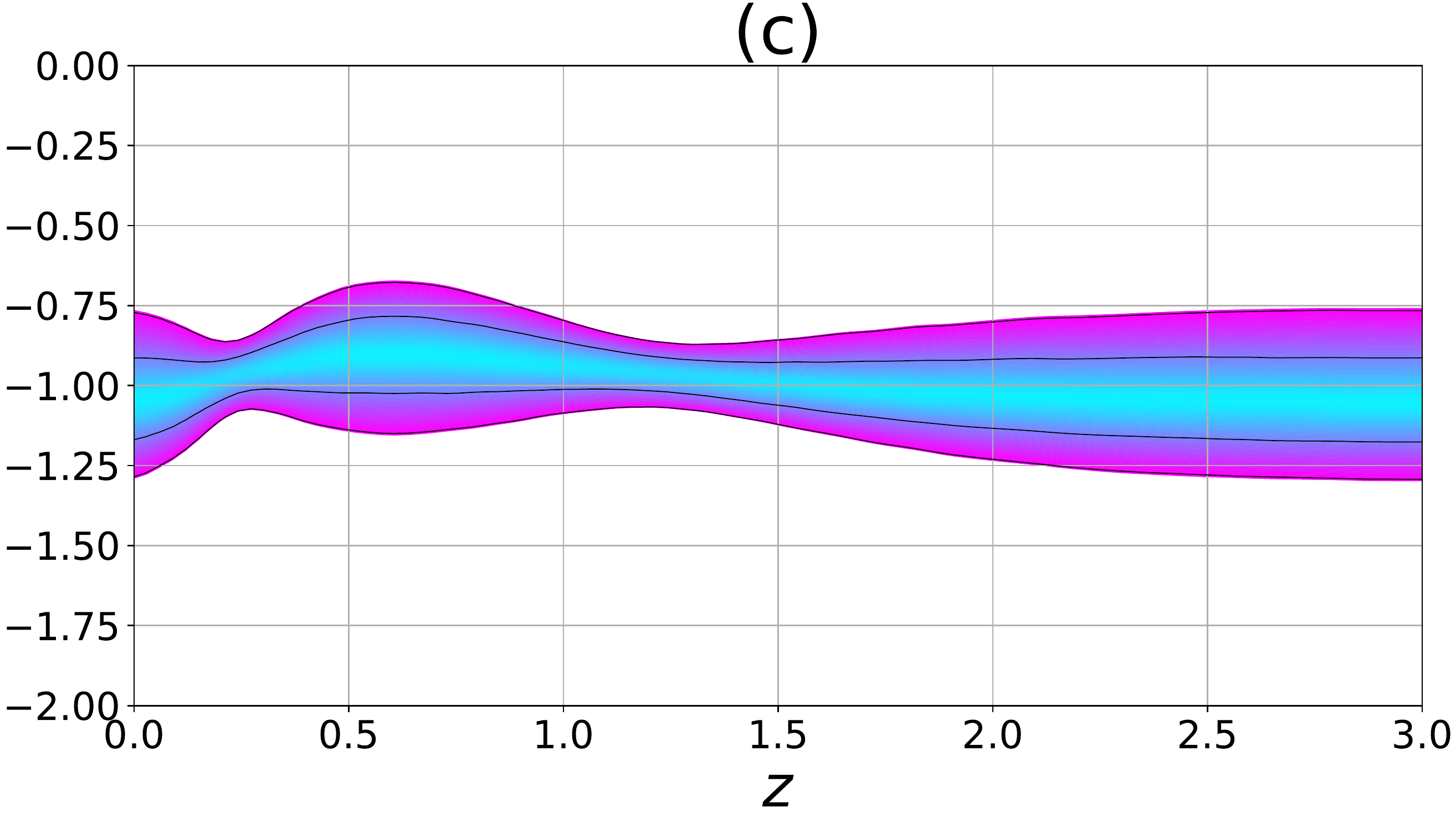}
        }
\subfigure{           \label{fig:d}
          \includegraphics[trim = 0mm 0mm 0mm 0mm, clip, width=4.2cm, height=3cm]{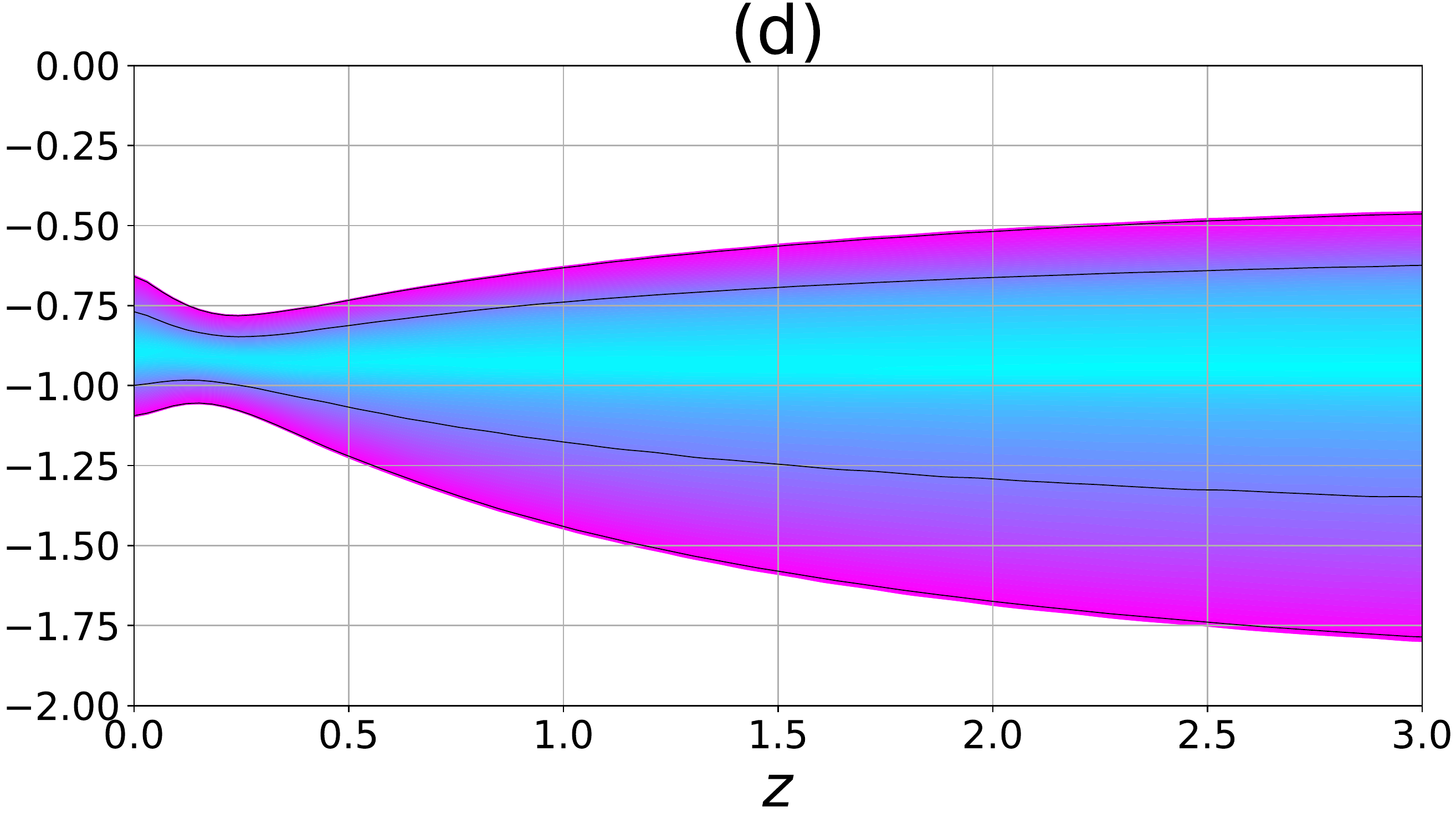}
        \includegraphics[trim = 0mm 0mm 0mm 0mm, clip, width=4.2cm, height=3cm]{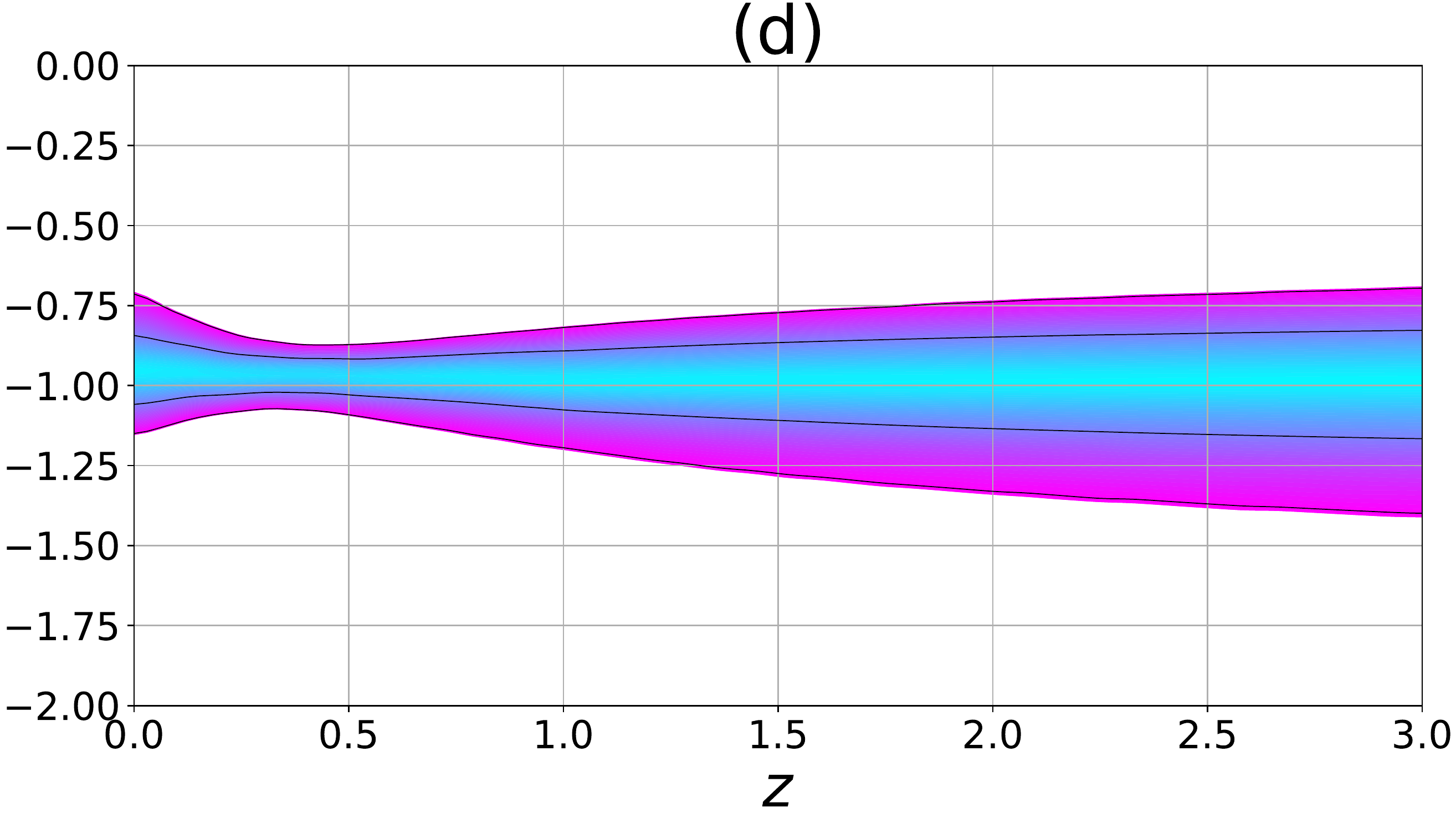}
        }
        
\subfigure{          \label{fig:e}
          \includegraphics[trim = 0mm 0mm 0mm 0mm, clip, width=4.2cm, height=3cm]{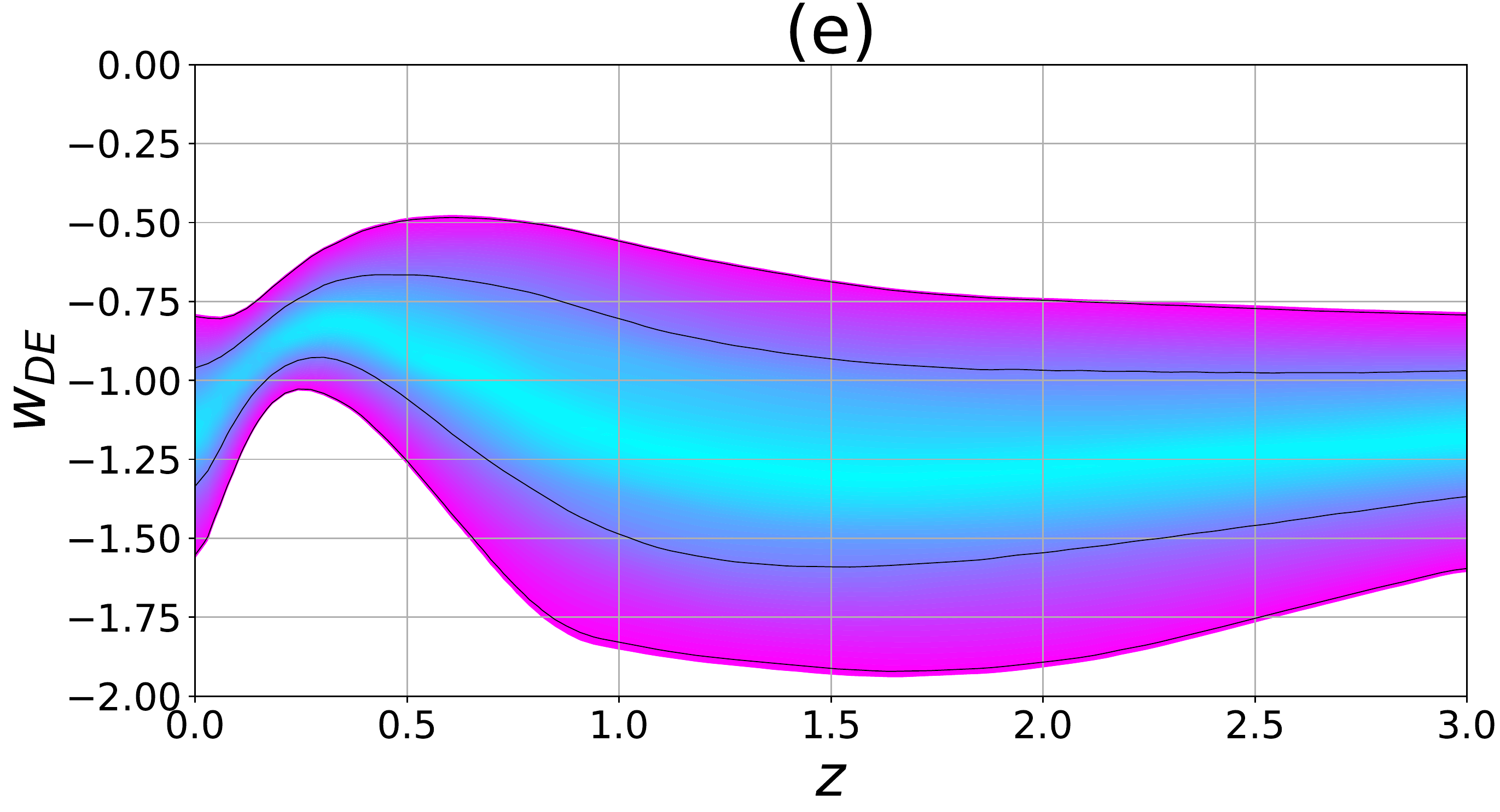}
          \includegraphics[trim = 0mm 0mm 0mm 0mm, clip, width=4.2cm, height=3cm]{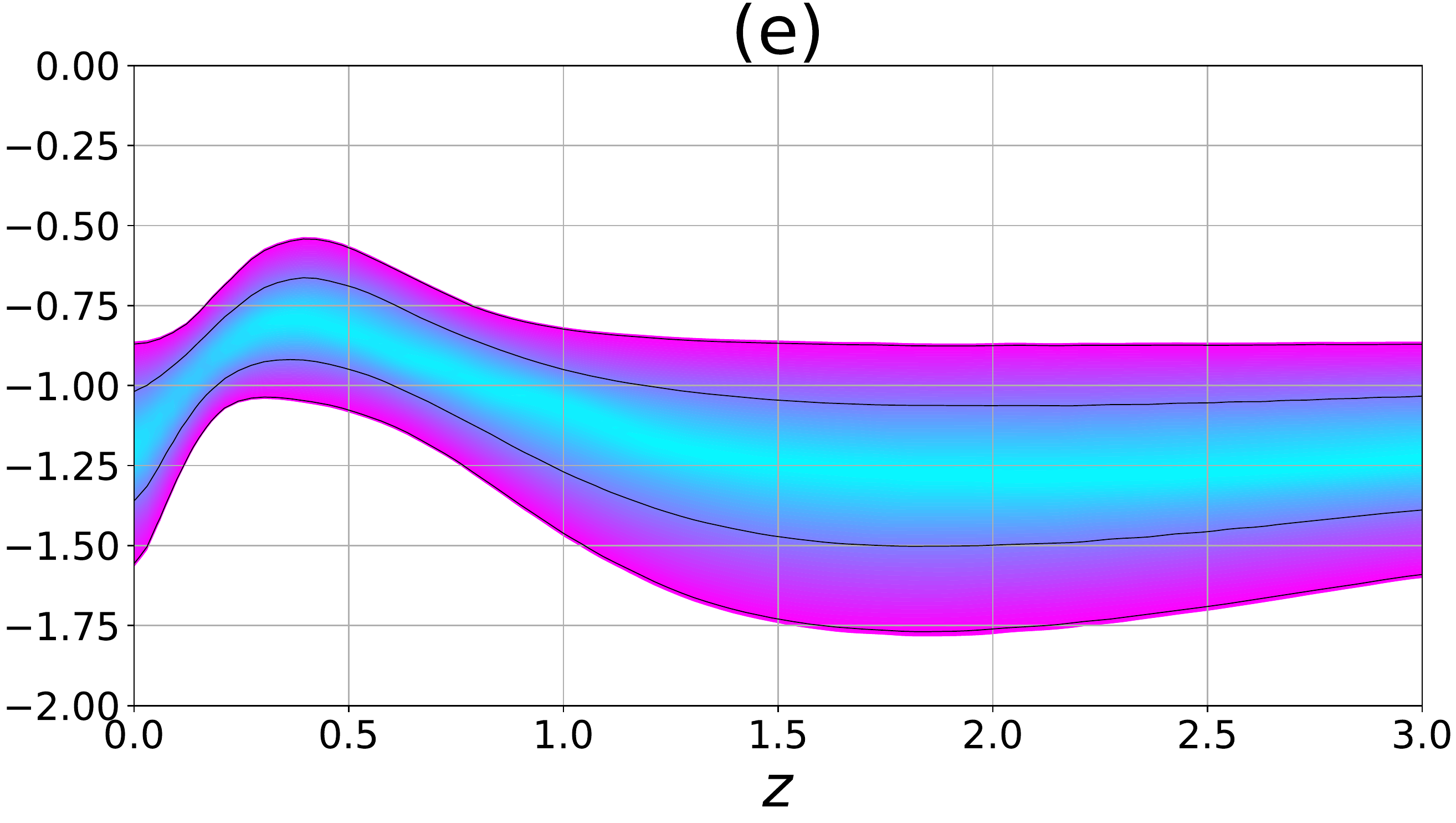}
        }
\subfigure{           \label{fig:f}
          \includegraphics[trim = 0mm 0mm 0mm 0mm, clip, width=4.2cm, height=3cm]{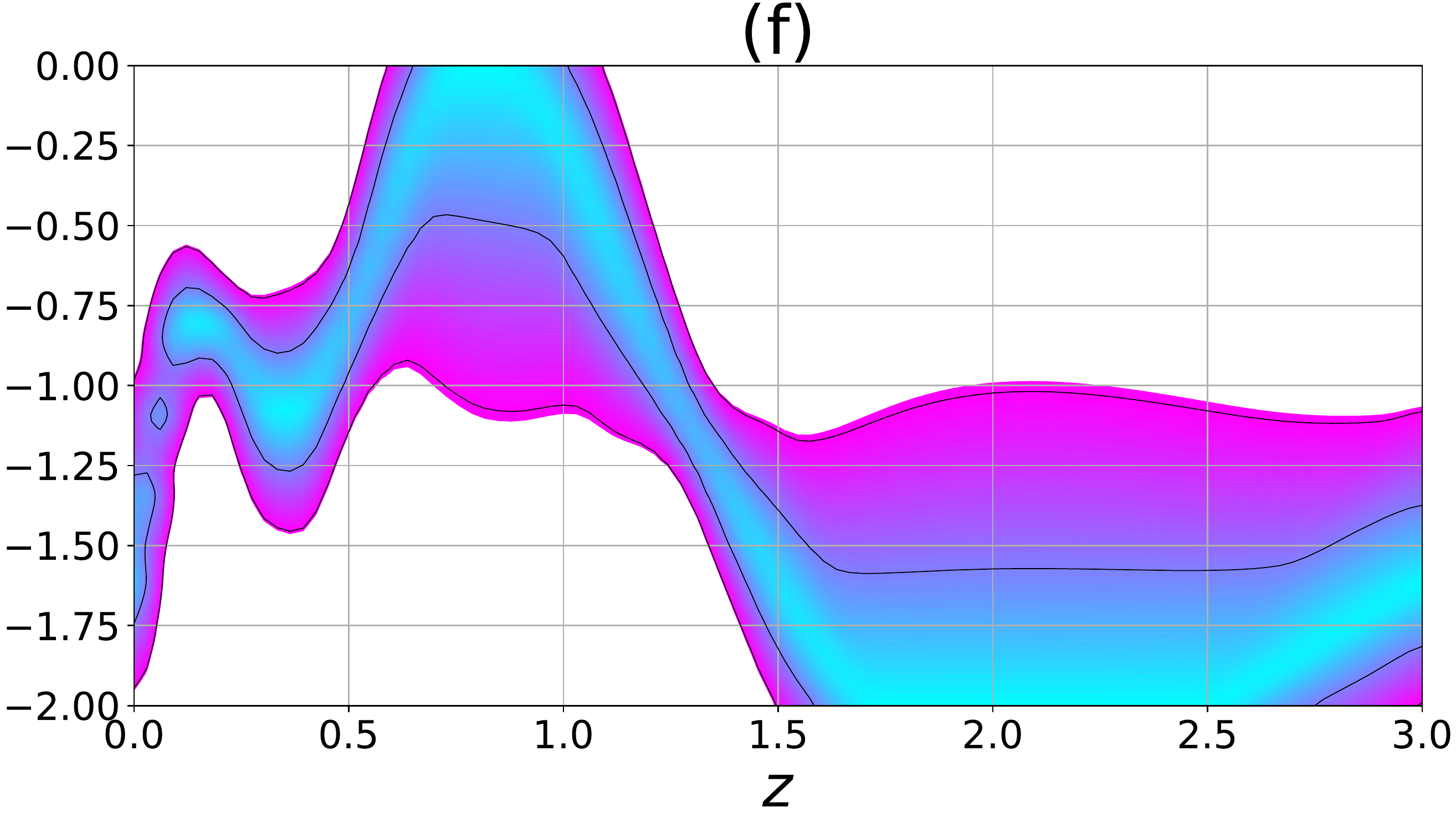}
          \includegraphics[trim = 0mm 0mm 0mm 0mm, clip, width=4.2cm, height=3cm]{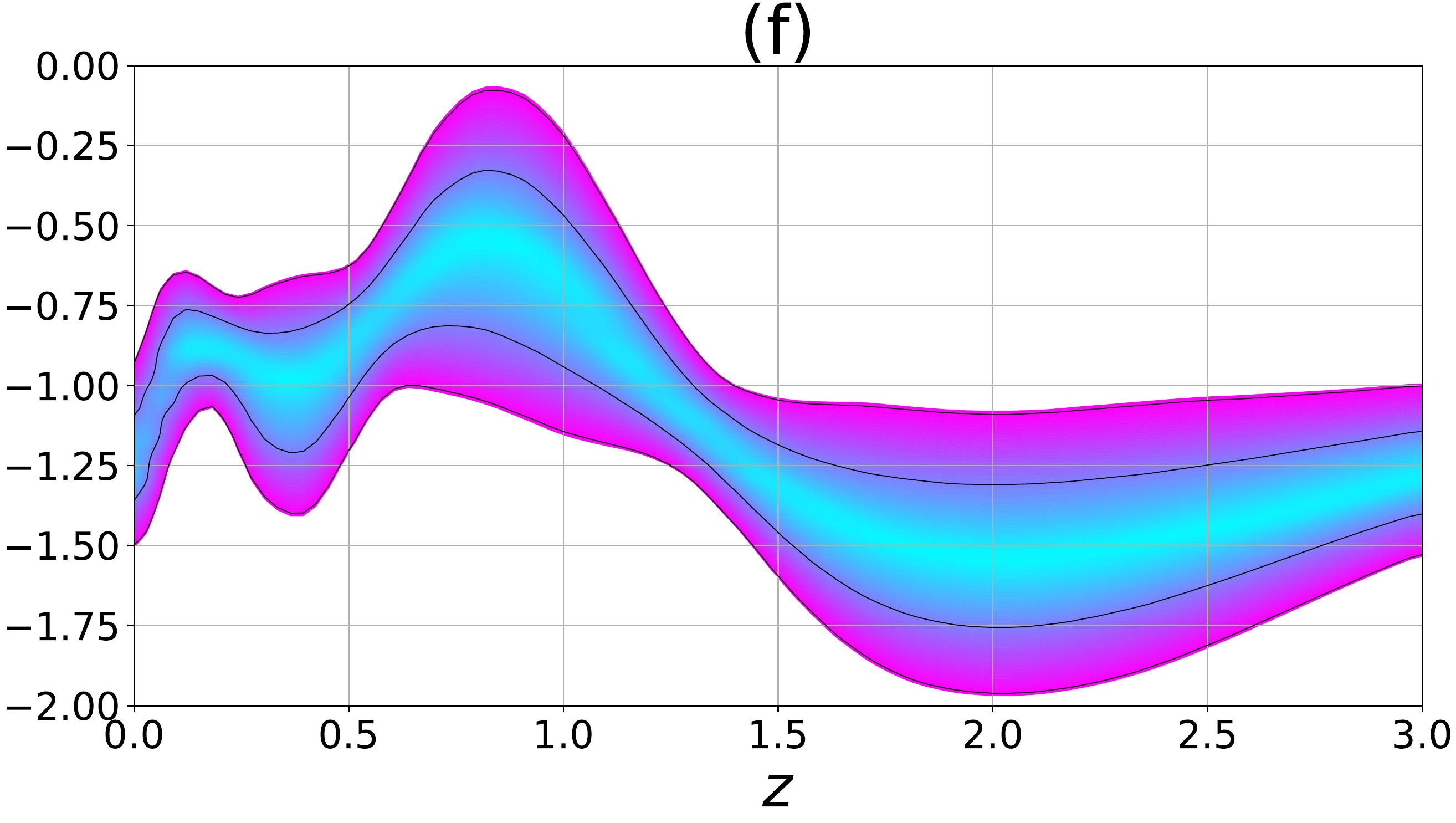}
        }
\subfigure{           \label{fig:g}
          \includegraphics[trim = 0mm 0mm 0mm 0mm, clip, width=4.2cm, height=3cm]{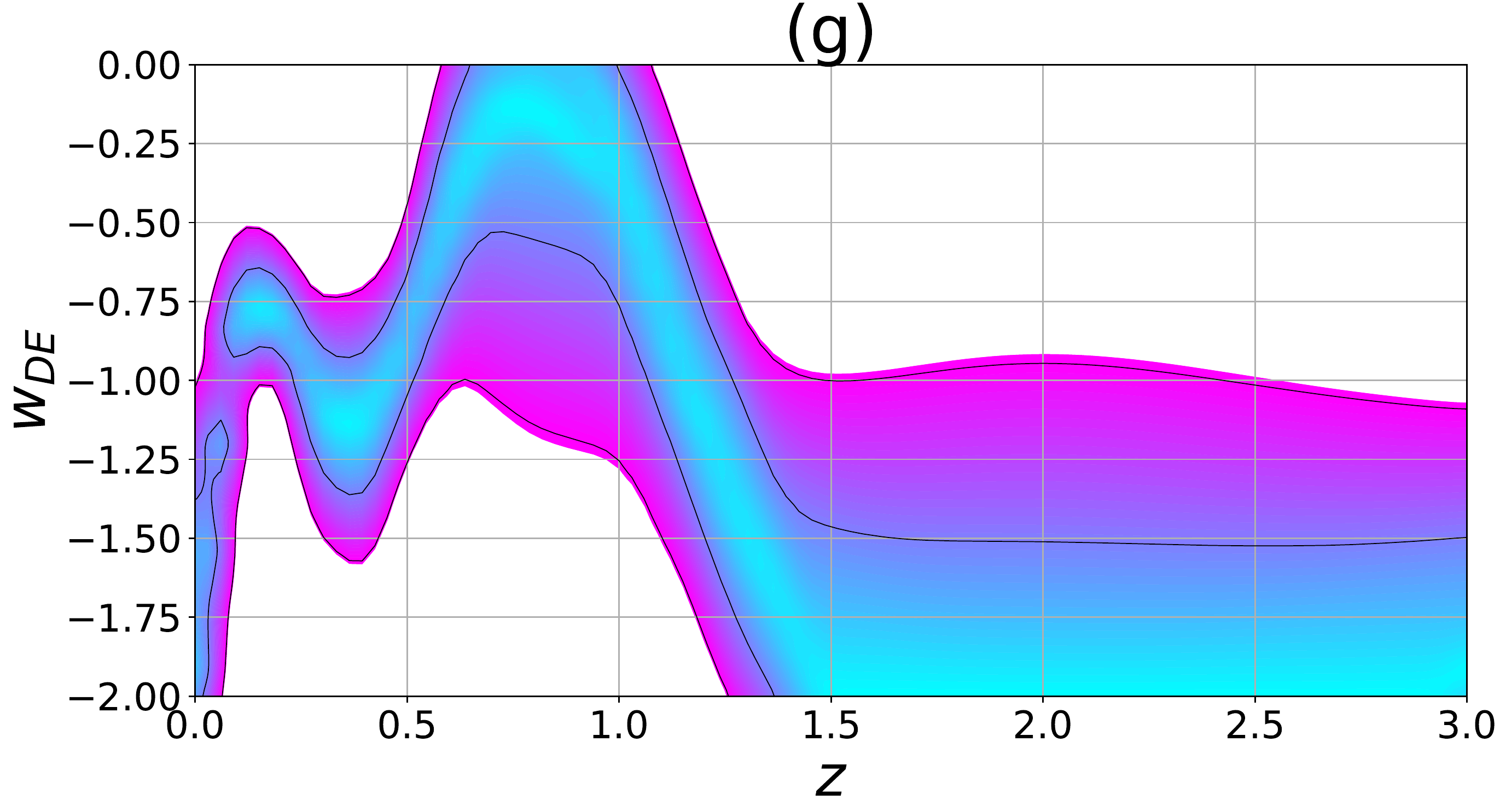}
          \includegraphics[trim = 0mm 0mm 0mm 0mm, clip, width=4.2cm, height=3cm]{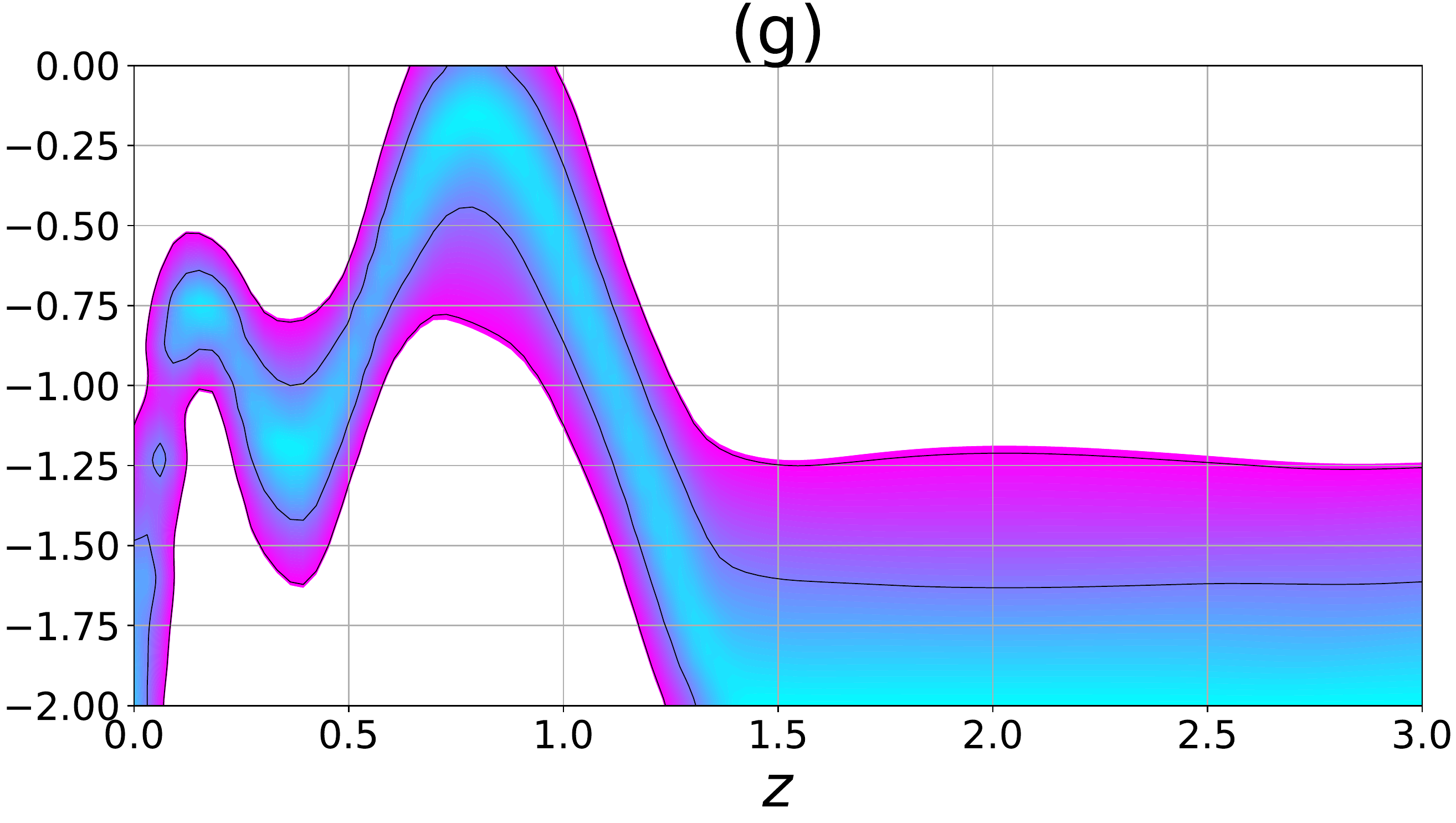}
        }  
\subfigure{           \label{fig:h}
          \includegraphics[trim = 0mm 0mm 0mm 0mm, clip, width=4.2cm, height=3cm]{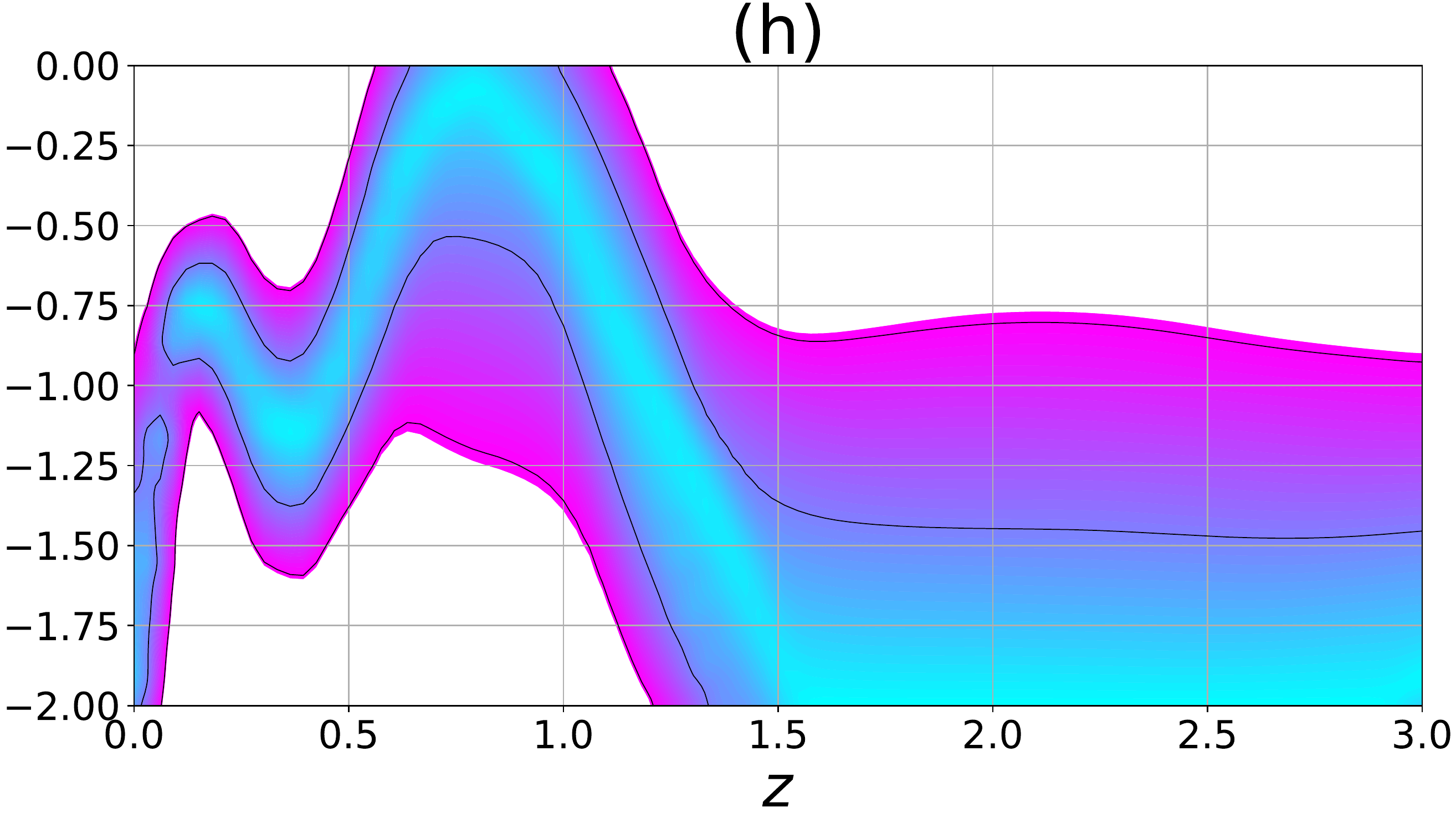}
          \includegraphics[trim = 0mm 0mm 0mm 0mm, clip, width=4.2cm, height=3cm]{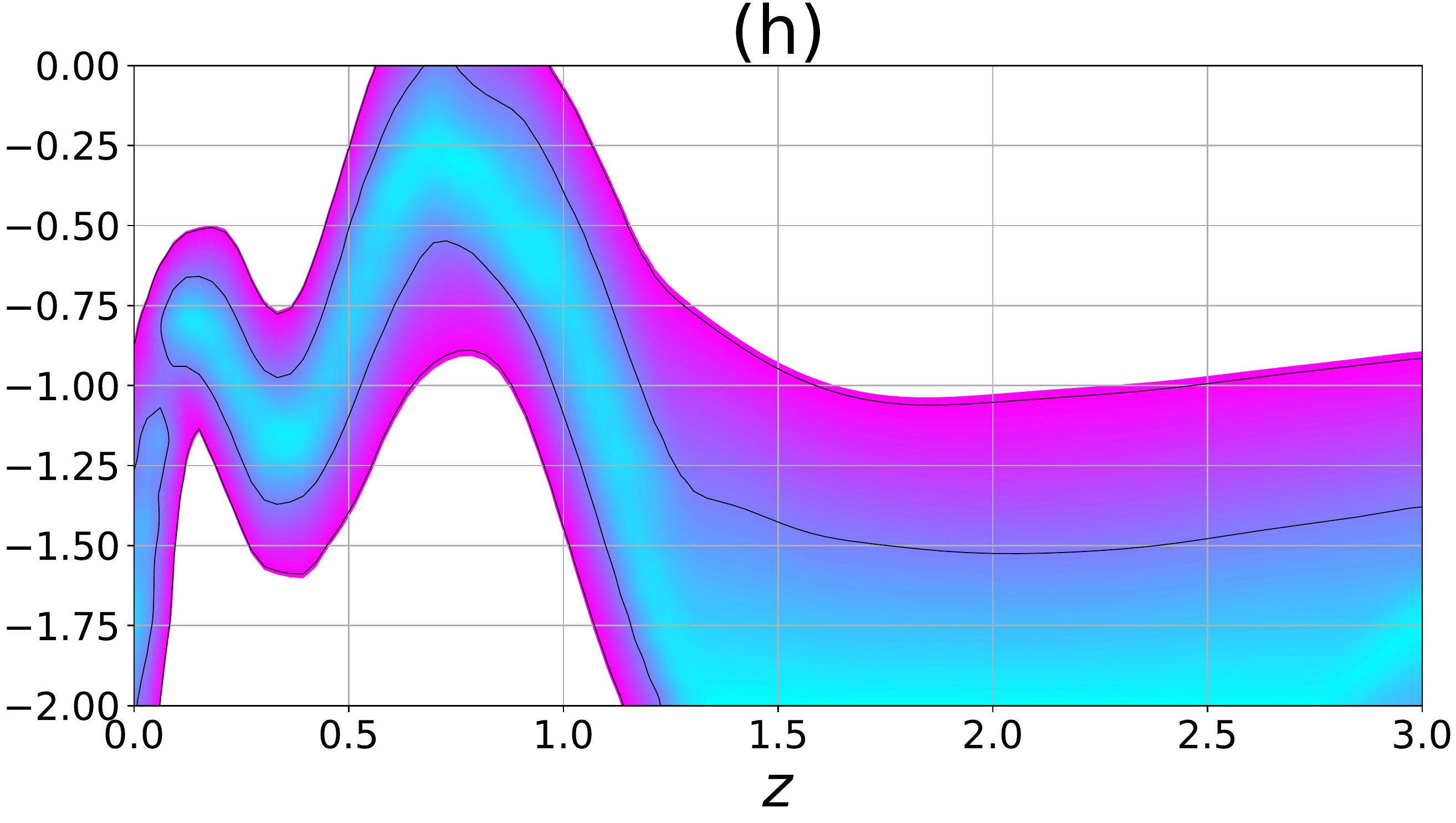}
        }
\caption{ 
These plots show the posterior probability ${\rm Pr}(w|z)$: the probability 
of $w$ as normalised in each slice of constant $z$, with colour scale in confidence interval values. 
The $1\sigma$ and $2\sigma$ confidence intervals are plotted as black lines.
Left panel for each set contains SN+BAO+HD datasets, while the right panel  
includes additionally PLK data.}
 \label{fig:posteriors}
\end{figure}

Table \ref{table} displays the mean and $1\sigma$ error values obtained during the analysis 
for the coefficients on each Fourier expansion. 
For each model the first row uses data from SNIa+BAO+HD, 
while the second row additionally includes PLK data.
The last three columns of the table contain statistical information
in order to provide an insight for the best model.
Notice that some of the best-fit values of the parameters are located right 
outside the $\Lambda$CDM model, within statistical significance. 
Of particular interest are the last three models where deviations
from the standard values $a_1=b_1=a_2=0$ are more noticeable, and hence
leading to a significant improvement on the likelihood. 
For instance model (f), with three extra parameters and fixed $w_0$,
has deviations from  $\Lambda$CDM  at about $2\sigma$ --according
to the signal-to-noise ratio in the fit alone--. 
Moreover, if $w_0$ is let to be a free parameter in model (g), 
the best-fit improves significantly
as a consequence of $w_0$ being different from the standard value $w_0= -2$
and deviations from $\Lambda$CDM increment up to $3\sigma$.
The inclusion of extra parameters improves the fit to the data (dash lines in 
Figure \ref{fig:delta}), however, also carries out 
a penalisation factor that affects directly the Akaike criteria (solid lines 
in Figure \ref{fig:delta}). 
That is, even though model (h) contains an extra parameter $b_2$, however, it
has no impact on improving the fit considerably. 
This is a consequence of $b_2$ being 
close to zero and hence providing no 
contribution to enhance the description of the data. 
Nevertheless, because of this increment of the number of parameters the
penalty term takes over and hence the AIC value raises back again.
We can go even further adding parameters, however as we have seen, 
the penalisation factor dominates 
and also more freedom brings more correlations between parameters and therefore
noisier reconstructions. 
Figure \ref{fig:posteriors} shows the posterior distribution (with $1\sigma$
and $2\sigma$ confidence levels) for the equation-of-state 
$w(z)$ given the set of MC chains for each description. 
As expected, adding parameters provides more structure than the cosmological 
constant. Let us have a look again at the last three models, where
the shape of $w(z)$ resembles a similar form already obtained in previous analyses, 
i.e. \cite{Vazquez:2012ce, Hee:2016nho, Zhao:2017cud}.
Throughout these reconstructions we notice the presence of two peaks, the major one
located at $z\sim 0.8$ and a small one at $z\sim 0.2$ (similar positions 
to the ones obtained in paper \cite{Zhao:2017cud}). 
It is also observed that at the present time ($z=0$) and high redshifts ($z>1$) 
 slightly favoured $w<-1$,  while at redshift $(z \sim 0.8)$ $w>-1$ is 
preferred, and hence the reconstructed $w(z)$ exhibits the crossing of the PDL several times. The crossing of the PDL plays a key role in 
identifying the correct dark energy model. 
If future surveys confirm its existence, single scalar field theories 
(with minimal assumptions) might be in serious problems as they cannot 
reproduce this essential feature, and therefore alternative models should be considered.
A key point to stress out is that the cosmological constant $w=-1$ 
lays down far outside of the $2\sigma$ region (outer solid black line), 
particularly at high redshift on the second plot of model (g) of Figure \ref{fig:posteriors}. 
The richness of this form is a consequence of releasing  some tensions between 
datasets, specially for the high-$z$ BAO. 

\begin{figure}
\includegraphics[trim = 0mm 0mm 0mm 0mm, clip, width=8cm, height=8cm]{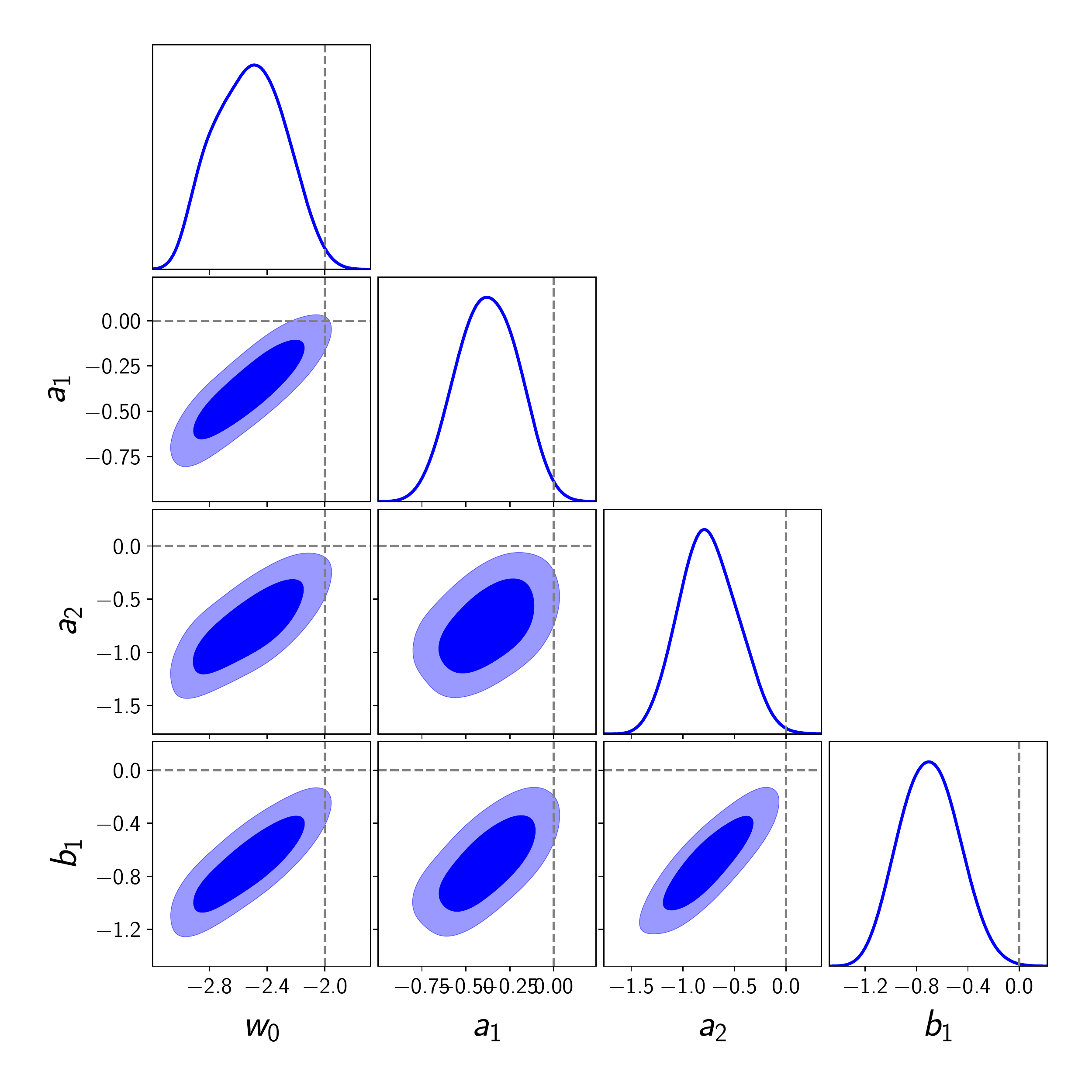}
\includegraphics[trim = 0mm 0mm 0mm 0mm, clip, width=6cm, height=8cm]{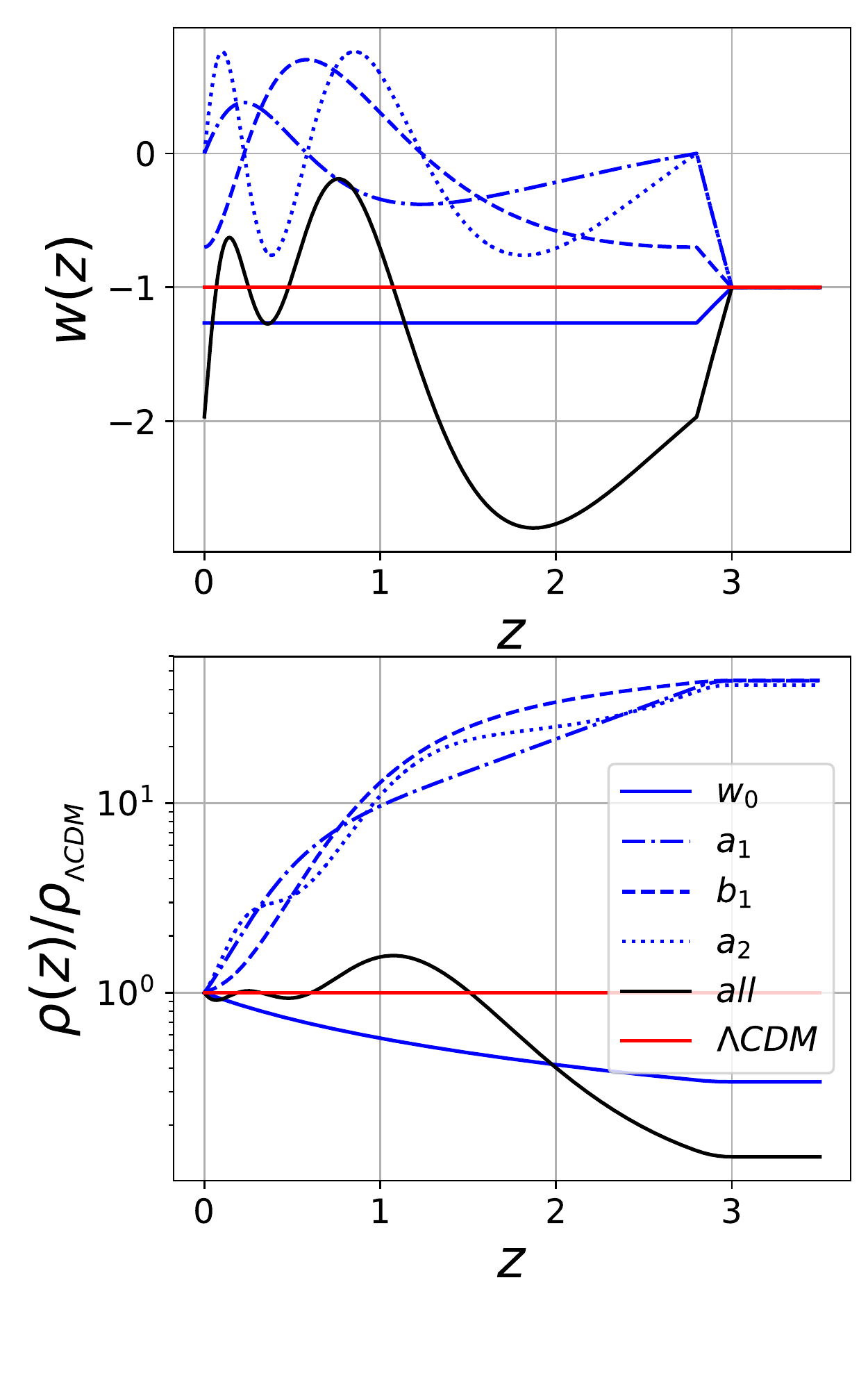}
\caption{ 
{\bf Left panel:} 1D and 2D probability posterior distributions of the parameters used in model (g) with the incorporation of PLK data. Vertical dashed lines correspond to the $\Lambda$CDM values.
{\bf Right panel:} $w(z)$ and $\rho(z)$ contributions from each term in the Fourier expansion. The best-fit values of the left panel were used in this plot.}
 \label{fig:model_g}
\end{figure}

Of all the models presented in Table \ref{table}, model (g) deviates
the most from the cosmological constant ($3.1\sigma$). 
Figure~\ref{fig:model_g} displays 1D and 2D marginalised posterior distributions 
for the parameters used to describe model (g). 
The vertical dashed lines, which correspond to the $\Lambda$CDM model, 
give an insight to the amount of deviation that each parameter presents,
in particular parameters $b_1$ and $a_2$ deviate the most from the standard values. 
Here we also notice some parameters are highly correlated. 
In a future work we would perform a dimensional reduction analysis (i.e. PCA)
which decreases the penalty factor, but still preserves a similar shape of $w(z)$ 
and hence $-2\ln \mathcal{L}$ keeps a similar value.
On the other hand, the right panel of Figure \ref{fig:model_g}
displays the $w(z)$ and $\rho(z)$ contributions from each term in the Fourier 
expansion. The $a_2$ parameter contribution looks very alike to the full 
Fourier expansion, and hence its importance in the reconstruction. 
Similarly the $b_1$ parameter enhances the amplitude of the major peak and 
provides contributions to the low-negative values of $w(z)$ and redshift today.

\section{Conclusions and Discussions}\label{section: Conclusions}

In this work were proposed Fourier series to describe the dynamical dark energy EoS parameter $w(z)$.
This approach reproduces the oscillating behaviour of $w(z)$ shown in reconstructions from observational data.
Also generalises previous dynamical dark energy proposals from trigonometrical functions and avoids some problems inherent of models based on the Taylor series, like the divergence at relative high redshift ($\sim 3$).
For several selected cases of the Fourier series, the parameters were constrained from data using a simpleMC and compared with other models such as $\Lambda$CDM and a particular case of Taylor series (CPL model). 
We noticed that as the number of parameters increase (i.e. terms of the series), more
correlations amongst them are created, and also an increment to the penalisation factor in the Akaike criteria. 
The Fourier series approach yields to a better fit to the data by more than $3\sigma$ in comparison to the cosmological constant $w=-1$, therefore having a clear preference for a dynamical dark energy behaviour. 
Moreover, if we compare the particular model (g) against the CPL parameterization, we found that model (g) is  preferred by more than $3\sigma$, also the difference in the Akaike criteria is $\Delta {\rm AIC_C}=-4.9$, which can be considered as a strong evidence against the CPL description.
With this analysis, in a model independent way, we are able to discriminate the cosmological constant and provide a better fit than the Taylor expansion -- in particular the CPL parameterization--. 
We have considered the  ${\rm AIC_C}$ to penalise the extra parameters introduced in our analysis. 
Even though this penalisation factor acts strongly, the criteria still tells us that a Fourier expansion provides a slightly better explanation of the data, specially to the inclusion of BAO at high redshift. 
This first analysis of considering $w(z)$ as a Fourier series, has its success in having a natural oscillatory behaviour and being compatible with model-independent reconstructions. 
More and better observational data are needed to test in more detail this proposal, however, it seems that the multiple crosses of the PDL are unavoidable, putting in conflict some simple models for dark energy
such as the cosmological constant, simple scalar fields, CPL and low-order Taylor series.

\section{Acknowledgements}
JAV acknowledges the support provided by FOSEC SEP-CONACYT Investigaci\'on B\'asica A1-S-21925, and DGAPA-PAPIIT  IA102219. 
DT acknowledge financial support from CONACYT Mexico postdoctoral fellowships.
The authors thank Josu\'e De-Santiago for the revision of the manuscript.

\end{document}